\documentclass[apj]{emulateapj}
\usepackage{psfig}
\begin{document}
\title{The Radio Emission, X-ray Emission, and Hydrodynamics of
  G328.4+0.2: A Comprehensive Analysis of a Luminous Pulsar Wind
  Nebula, its Neutron Star, and the Progenitor Supernova Explosion}

\author{Joseph D. Gelfand} 
\affil{Harvard-Smithsonian Center for Astrophysics, Cambridge, MA 02138} 
\email{jgelfand@cfa.harvard.edu}
\and 
\author{B. M. Gaensler\footnote{Alfred P. Sloan Research
  Fellow, Australian Research Council Federation Fellow}}
\affil{Harvard-Smithsonian Center for Astrophysics, Cambridge, MA 02138} 
\affil{School of Physics, The University of Sydney, NSW 2006, Australia}
\author{Patrick O. Slane}
\affil{Harvard-Smithsonian Center for Astrophysics, Cambridge, MA 02138}
\author{Daniel J. Patnaude}
\affil{Harvard-Smithsonian Center for Astrophysics, Cambridge, MA 02138}
\author{John P. Hughes}
\affil{Department of Physics and Astronomy, Rutgers University, Piscataway, 
NJ 08854-8019}
\author{Fernando Camilo} 
\affil{Columbia Astrophysics Lab, Columbia University, New York, NY 10027}

\begin{abstract}
We present new observational results obtained for the Galactic
non-thermal radio source G328.4+0.2 to determine both if this source
is a pulsar wind nebula or supernova remnant, and in either case, the
physical properties of this source.  Using X-ray data obtained by {\it
XMM}, we confirm that the X-ray emission from this source is heavily
absorbed and has a spectrum best fit by a power law model of photon
index $\Gamma=2$ with no evidence for a thermal component, the X-ray
emission from G328.4+0.2 comes from a region significantly smaller
than the radio emission, and that the X-ray and radio emission are
significantly offset from each other.  We also present the results of
a new high resolution ($7^{\prime \prime}$) 1.4~GHz image of
G328.4+0.2 obtained using the Australia Telescope Compact Array, and a
deep search for radio pulsations using the Parkes Radio Telescope.  By
comparing this 1.4~GHz image with a similar resolution image at
4.8~GHz, we find that the radio emission has a flat spectrum ($\alpha
\approx 0$; $S_{\nu} \propto \nu^\alpha$), though some areas of the
eastern edge of G328.4+0.2 have a steeper radio spectral index of
$\alpha \sim -0.3$.  Additionally, we searched without success for a
central radio pulsar, and obtain a luminosity limit of $L_{1400} < \la
30$\,mJy\,kpc$^2$, assuming a distance of 17\,kpc.  In light of these
observational results, we test if G328.4+0.2 is a pulsar wind nebula
(PWN) or a large PWN inside a supernova remnant (SNR) using a simple
hydrodynamic model for the evolution of a PWN inside a SNR.  As a
result of this analysis, we conclude that G328.4+0.2 is a young
($\la10000$~years old) pulsar wind nebula formed by a low magnetic
field ($\la10^{12}$~G) neutron star born spinning rapidly ($\la10$~ms)
expanding into an undetected SNR formed by an energetic
($\ga10^{51}$~ergs), low ejecta mass ($M_{\rm ej}\la 5M_{\odot}$)
supernova explosion which occurred in a low density
($n\sim0.03$~cm$^{-3}$) environment.  If correct, the low magnetic
field and fast initial spin period of this neutron star poses problems
for models of magnetar formation which require fast initial periods.
\end{abstract}
\keywords{stars: neutron, stars: pulsars: general, ISM: supernova
  remnants, radio continuum: ISM, X-rays: individual}

\section{Introduction}
\label{intro}

Stars with initial masses between $\sim$9 and 25 M$_\odot$ are
expected to end their lives in a giant supernova (SN) explosion during
which neutron stars are created.  The fast moving ejecta from the SN
create a supernova remnant (SNR), while the particle wind produced by
the neutron star as it loses rotational energy inflates a pulsar wind
nebula (PWN; \citeauthor{gaensler06} \citeyear{gaensler06}).
Initially, the PWN is inside the SNR -- and when the PWN is detected
inside the SNR the system is called a ``composite'' SNR
\citep{helfand87}.  The evolution of the central neutron star and the
outer SNR affect the PWN, and as a result the PWN goes through several
evolutionary phases while it is inside the SNR.  This evolution is
determined by the physical properties of the neutron star
(specifically the initial period $P_0$, the braking index
$p$\footnote{The braking index is defined as $\dot{\Omega} \propto
  \Omega^p$, where $\Omega$ is the angular velocity of the neutron
  star's surface.}, and the strength of the dipole component to the
surface magnetic field $B_{\rm ns}$), the SN explosion (explosion
energy $E_{\rm sn}$ and ejecta mass $M_{\rm ej}$), and the surrounding
medium (ambient number density $n$).  As a result, by measuring the
properties of the PWN inside a SNR at a given time one is able to
constrain these physical parameters which allows one to study the
mechanisms behind both core-collapse SNe and massive star evolution.

With this in mind, we present results of new radio observations of
this source with the Australia Telescope Compact Array (ATCA), as well
as a new X-ray (X-ray Multi-mirror Mission; {\it XMM}) observation of
G328.4+0.2 (MSH~15-5{\it 7}; \citeauthor{mills61} \citeyear{mills61}).
This source is a distant ($d \geq 17.4 \pm 0.9$~kpc;
\citeauthor{gaensler00} \citeyear{gaensler00}), radio bright (flux
density $S_{\nu}=14.3 \pm 0.1$~Jy at $\nu=$1.4~GHz), polarized,
extended (diameter $D \simeq 5.0^{\prime}$) radio source with a
relatively flat spectral index ($\alpha \simeq -0.12\pm0.03$ where
$S_{\nu}\propto\nu^\alpha$; \citeauthor{gaensler00}
\citeyear{gaensler00}).  Based on these radio properties, and the
discovery of non-thermal X-ray emission from this source by {\it ASCA}
\citep{hughes00}, this source was classified as a PWN -- the largest 
and most radio-luminous PWN in the Galaxy. In this interpretation, the
expectation is that G328.4+0.2 is $\sim7000$ years old and powered by
an extremely energetic neutron star \citep{gaensler00}.  However,
follow up radio polarimetry work \citep{johnston04} implied that
G328.4+0.2 is an older composite SNR in which the PWN is just a small
fraction of the total volume, and as a result is powered by a
significantly less energetic neutron star then argued by
\citet{gaensler00}.  In this paper, we analyze new observations of
this source in order to determine the age of G328.4+0.2, the
energetics of the neutron star and the progenitor SN, and the density
of its environment.

In \S\ref{data} we present new X-ray and radio observations of
G328.4+0.2.  In \S\ref{theory}, we first discuss the expected
evolutionary sequence for PWNe inside SNRs (\S\ref{evolution}), making
general comments regarding the expected observational signature of
each phase.  In \S\ref{obsres} use the observational results presented
in \S \ref{data} to draw some initial conclusions about the nature of
G328.4+0.2.  In \S\ref{model}, we present a simple hydrodynamical
model for the evolution of a PWN inside a SNR, which we apply to
G328.4+0.2 assuming it is a composite SNR (\S\ref{g328comp}) or a PWN
(\S\ref{g328pwn}).  Finally, in \S\ref{conclusion} we summarize our
results. 

\section{Observations}
\label{data}

In this Section, we present the data gathered in a {\it XMM}
observation (\S\ref{xray}), a 1.4~GHz ATCA observation of G328.4+0.2
(\S\ref{radio}), and a search for a radio pulsar in this source
(\S\ref{pulsar}).

\subsection {X-ray Observations}
\label{xray}

On 2003 March 9--10, G328.4+0.2 was observed for $\sim$50~ks by {\it XMM}.
During this observation, the {\sc pn} camera was operated in Small
Window Mode, and the {\sc Mos1} and {\sc Mos2} camera were operated in
Full Frame Mode.  The ``Thick'' optical filter was used due to the
presence of numerous bright stars in the field-of-view of G328.4+0.2.  The
data were reduced with the software package {\sc xmm-sas v 6.0.0} with
calibration files current through {\sc XMM-CCF-REL-174}, using the
standard procedure for reducing {\it XMM} data outlined in the {\it
  XMM-Newton ABC Guide}\footnote{Available at
http://heasarc.gsfc.nasa.gov/docs/xmm/abc/} and the {\it Birmingham 
XMM Guide}\footnote{Available at http://www.sr.bham.ac.uk/xmm2/guide.html}.  

\subsubsection{Image Analysis}
\label{image}

A vignetting corrected 0.2--12~keV image from the {\sc Mos1} and {\sc Mos2} 
instruments\footnote{We did not use the {\sc pn} data due to the
  substantially larger pixel size of this instrument.}, is shown in
Fig.~\ref{xrayimg}, in which we observe three spatial components to the
X-ray emission: a bright, compact feature located along the SW edge of
the X-ray emission (``Clump 1''), a fainter, slightly extended feature
located NE of the compact feature described above (``Clump 2''), and
extended diffuse emission, roughly $1^{\prime}$ in diameter,
surrounding the two features described above (``Diffuse'').  From the
``Clump 1'' region we detected $120\pm12$ counts above the background
between 0.5--10 keV in the {\sc Mos1} detector and  $136\pm12$ in the
{\sc Mos2} detector, from the ``Clump 2'' region we detected $66\pm9$
counts in both the {\sc Mos1} and {\sc Mos2} detectors, and from the
``Diffuse'' regions we detected $360\pm20$ and $380\pm20$ counts from
the {\sc Mos1} and {\sc Mos2} detectors, respectively.

We determined the spatial properties of these components using the 
{\sc Sherpa} modeling software package \citep{sherpa}.  Due to the low
number of counts per pixel, we used the {\it simplex} fitting method
and minimized the {\it cash} statistic \citep{cash79}.  We attempted
to model Clump~1 and Clump~2 as circular 2D Gaussians, elliptical 2D
Gaussians, or circular 2D Lorentzians, the latter of which is a good
model for {\it XMM}'s point spread function (PSF; \citeauthor{xmmpsf}
\citeyear{xmmpsf}).  We attempted to model the Diffuse region as a
circular or elliptical 2D Gaussian, and assumed a constant background.
We fit the observed image to all model combinations of
Clump~1, Clump~2, and Diffuse (attempts to eliminate one of these
components resulted in significantly worse fits), and the 
best fit was obtained for a model in which Clump~1 is a 2D~Lorentzian
while Clump~2 and Diffuse are elliptical 2D~Gaussians.  The fit
parameters for this model are given in Table \ref{sherpares}, and the
model and residuals are shown in Fig. \ref{xrayimg}, and from this
conclude that the emission from Clump~1 is consistent with the PSF of
{\it XMM}.  Additionally, from this fit we estimate that Clump~1 and
Clump~2 are separated by $\sim10\arcsec$, while the centers of Clump~1
and Diffuse are separated by $\sim15\arcsec$.

\subsubsection{Spectral Results}
\label{spectra}

\begin{figure}
\begin{center}
\includegraphics[width=4.5in,angle=0]{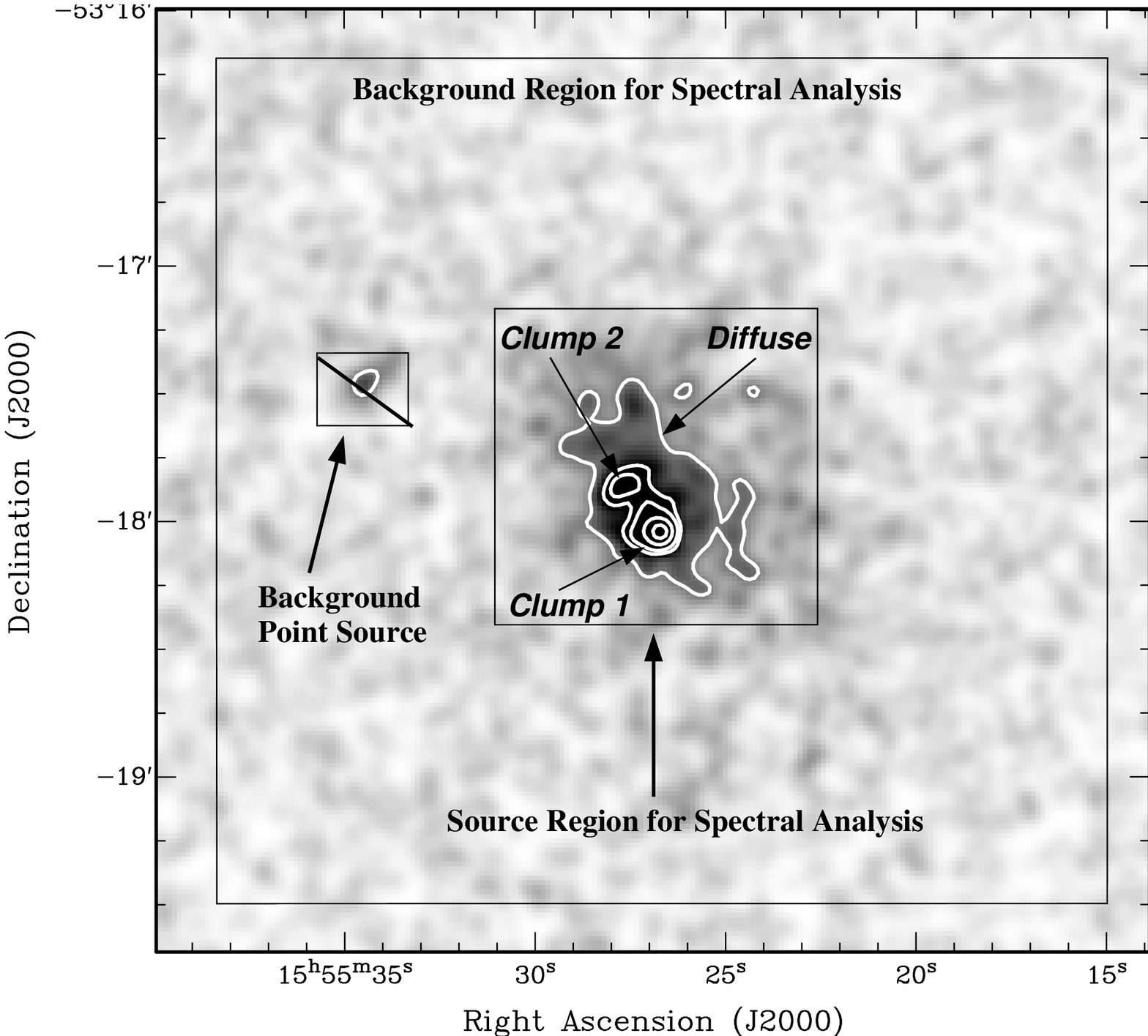}
\includegraphics[width=4.5in,angle=0]{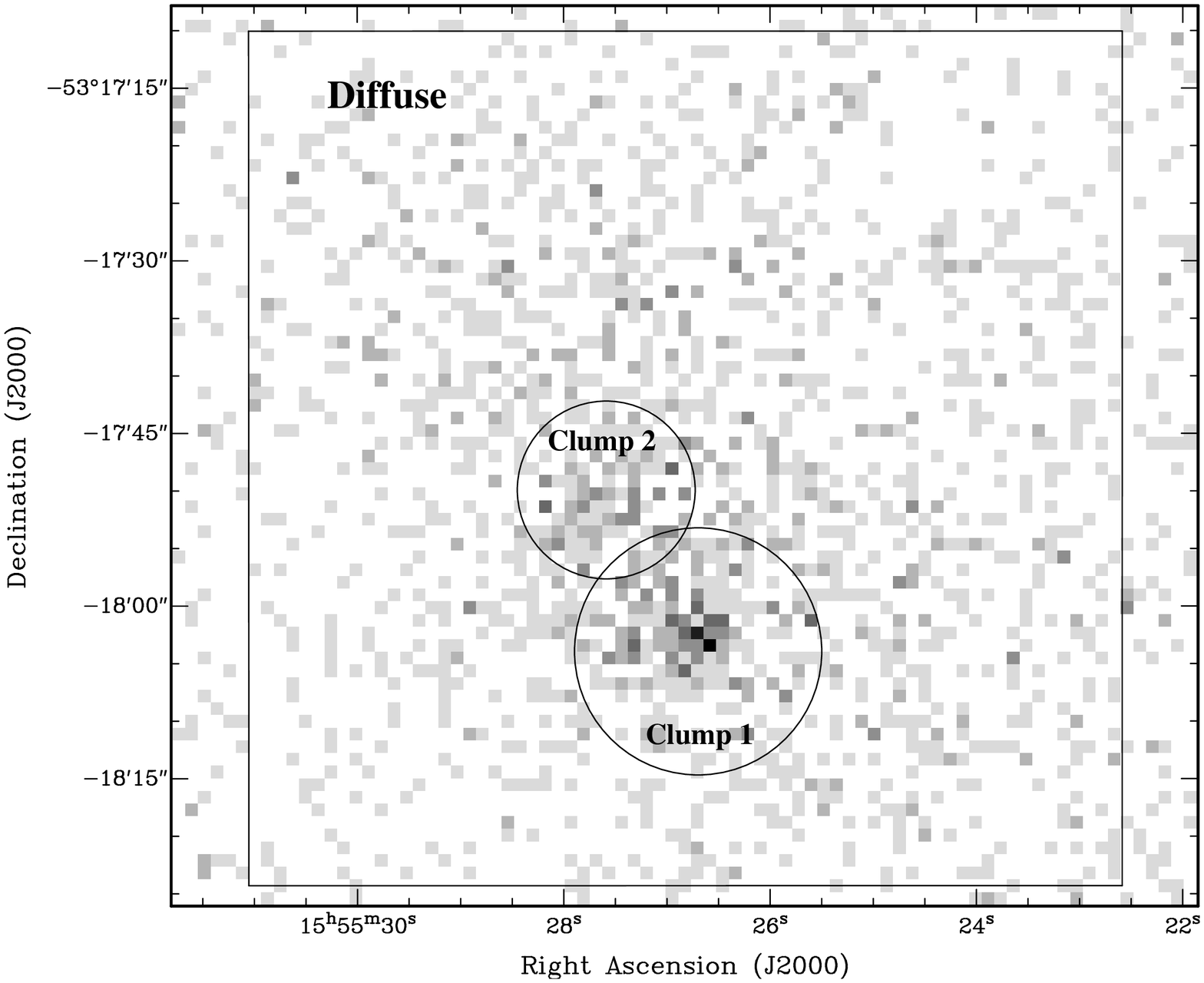}
\end{center}
\figcaption{{\it Top:} Exposure normalized, vignette corrected {\sc
  mos1 + mos2} image of G328.4+0.2 smoothed by a $5^{\prime
  \prime}$~Gaussian.  The white contours indicate 20, 40, 50, 70, and
  90\% of the peak X-ray flux in the smoothed image, while the boxes
  indicate the background and source regions used for the spectral
  analysis described in \S\ref{spectra}.  The background point source
  labeled in the image was excluded from the background region.
  Additionally, the labels in this plot point to the morphological
  features discussed in \S\ref{image}. {\it Bottom:} Unsmoothed
  normalized, vignette corrected {\sc mos1} and {mos2} image of
  G328.4+0.2 overlaid with the regions used for the Hardness Ratio
  analysis discussed in \S\ref{spectra}.\label{xrayimg}}      
\end{figure}

In generating the spectra, only events with {\sc flag = 0} were used.
Additionally, the event files were screened for background flares by
binning the 10--15~keV light curve of each instrument by 50~s and then
recursively flagging all bins with a count rate $>3\sigma$ above the
average.  This procedure removed 3.0~ks, 2.3~ks, and 0.6~ks of data
from the {\sc Mos1}, {\sc Mos2}, and {\sc pn} detectors, respectively.
Spectra were extracted for the regions shown in Fig. \ref{xrayimg},
and the resulting spectra were binned into a minimum of 25 counts per
channel and modeled using {\sc Xspec v12.2.0}.  The background regions
used are also shown in Fig.~\ref{xrayimg}.

\begin{figure}
\begin{center}
\includegraphics[width=7.5in,angle=0]{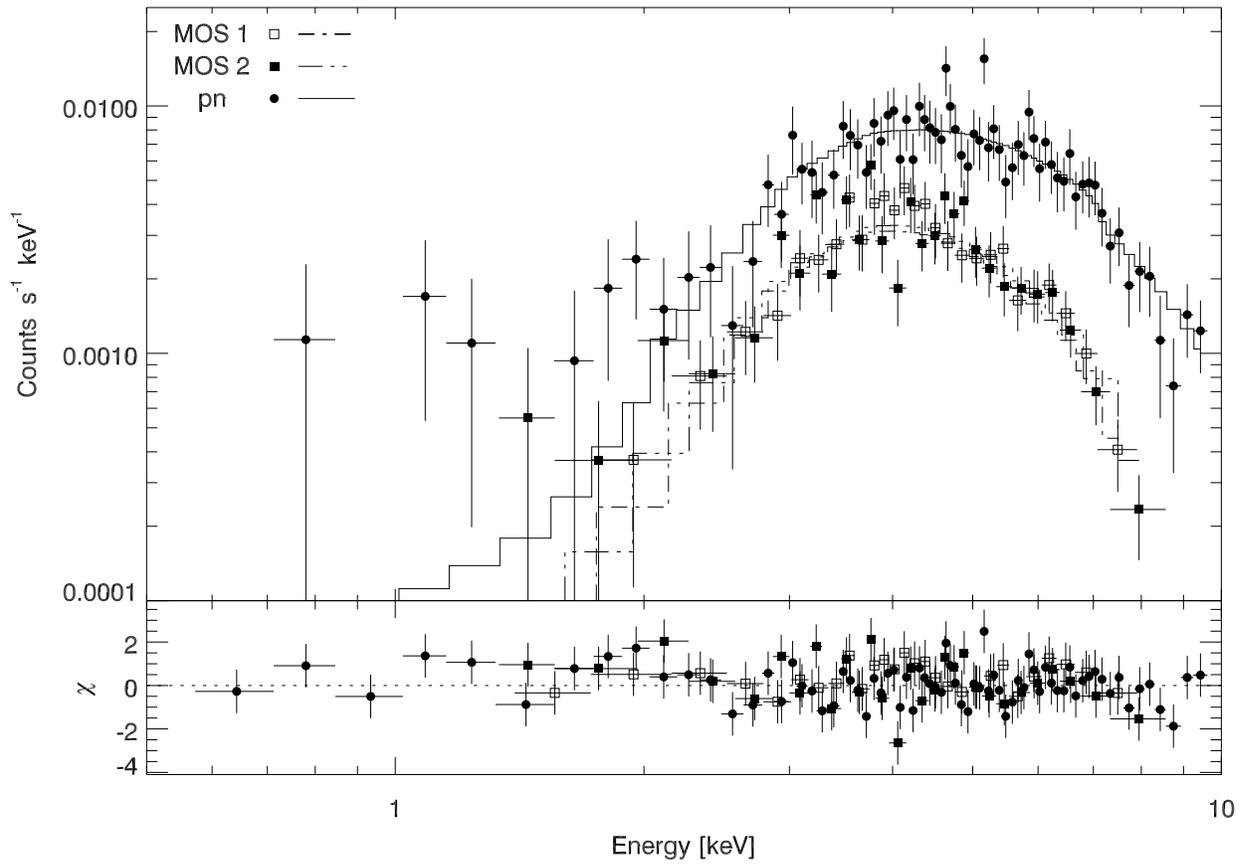}
\end{center}
\figcaption{{\sc Mos1}, {\sc Mos2}, and {\sc pn} spectrum of
  G328.4+0.2 overlaid with the absorbed power-law model whose
  parameters are given in Table \ref{specres}.\label{compspec}}  
\end{figure}

To determine the composite spectra of G328.4+0.2, we jointly fit the
spectrum obtained by the {\sc Mos1}, {\sc Mos2}, and {\sc pn}
detectors -- shown in Fig. \ref{compspec}.  The background-subtracted
observed 0.5--10~keV count rate of G328.4+0.2 was $0.012\pm0.001$ counts
s$^{-1}$ in the the {\sc Mos1} detector ($555\pm29$~counts),
$0.012\pm0.001$ counts s$^{-1}$ in the {\sc Mos2} detector
($580\pm30$~counts), and $0.045\pm0.001$ counts s$^{-1}$ in the {\sc
  pn} detector\footnote{The {\sc pn} detector count rate does not
  account for 29\% dead time since this instrument was operated in
  Small Window Mode.} ($1576\pm63$~counts).  We fit the spectra to seven
different models separately -- a power-law, a blackbody,
bremsstrahlung, a Raymond-Smith plasma, and a power-law plus one of
these three thermal models -- all attenuated for interstellar
absorption.  Only the single-component models produced reasonable fits
(reduced $\chi^2 \sim 1$), and the fitted parameters are presented in
Table \ref{specres}.  Both the blackbody ($kT_{\rm BB} \sim 1.7$~keV,
$T_{\rm BB}\sim20$~MK) and the bremsstrahlung models ($kT \sim 9$~keV)
require unrealistically high temperatures, especially if the X-rays
are from a PWN as argued by \citet{hughes00}.  The derived parameters
for the power law model are similar to that observed in other PWN
(e.g. \citeauthor{gotthelf03} \citeyear{gotthelf03}), and agree well
with the results obtained for G328.4+0.2 by \citet{hughes00}.  There
is no evidence for thermal X-ray emission, which one would expect from
a SNR, in this source.

Since the individual regions discussed in \S\ref{image} did not have
enough counts for spectral fitting, we measured their hardness ratio
({\sc HR}), defined as:
\begin{eqnarray}
\label{hrdef}
{HR} & = & \frac{H-S}{H+S}
\end{eqnarray}
where $H$ is the number of counts in the Hard (higher energy) band and
$S$ is the number of counts in the Soft (lower energy) band, of the
regions discussed in \S \ref{image} to determine if there were any
spatial variations in the X-ray spectrum of G328.4+0.2.  Using the
X-ray spectrum as a guide, we calculate HR with $H$ as the number of
counts between 4 and 8~keV and $S$ as the number of counts between 2
and 4~keV.  Since the pixels on the {\sc pn} detector are sufficiently
large that it is not possible to separate the emission from these
regions, we only use data from the {\sc mos1} and {\sc mos2}
detectors.  The calculated, background subtracted {\sc HR} of
G328.4+0.2 is $0.25\pm0.04$, of the Clump~1 region is $0.29\pm0.07$,
of the Clump~2 region is $0.31\pm0.11$, and of the diffuse region is
$0.24\pm0.05$ (1$\sigma$ errors).  As a result, we conclude that there
is no significant change in the X-ray spectrum of G328.4+0.2 between
these features.

\subsubsection{Timing Results}
\label{timing}

A clear signature for the presence of a neutron star would be the
detection of X-ray pulsations in the emission from G328.4+0.2.  We
searched for this using the $Z_n^2$ test defined by
\citet{buccheri83}, where $n$ is the harmonic number of periodic
signal, for $n=1,2,3,4$.  The maximum frequency searched was $\nu_{\rm
max}=1/(2n\Delta t)$, the minimum frequency searched was $1/20$~Hz,
and the frequency step was $1/4t_{\rm obs}$ ($5\times10^{-6}$~Hz,
oversampling the Nyquist rate by a factor of 2), where $\Delta t$ is
the time resolution of the dataset (5.7~ms) and $t_{\rm obs}$ is the
length of the observation.  We only used events from the {\sc pn}
instrument (5.7~ms since it was operated in Small Window Mode) due to
the poor time resolution (2.6s) of the {\sc mos} data.  Additionally,
we only used events from the Clump~1 region because only only emission
from the central NS should be pulsed and as the brightest X-ray region
of G328.4+0.2, this region is the most probable location of any
neutron star. Unfortunately, due to the large pixel size of the {\sc
pn} instrument this region is contaminated by emission from Clump~2
and the Diffuse region.  The event times from the resultant event list
were barycentered to the Solar System reference frame, and we searched
for a periodicity over multiple energy ranges in order to maximize the
sensitivity of our search.  The most significant period detected was
in the dataset which only included photons between 5 and 10~keV (159
photons), in which for $n=4$ a signal with period $P=336$~ms had
$Z_3^2=49.5$ in 2074786 independent trials for this value of $n$ and
energy range, which has a 12\% chance of being a false positive, a
$<2\sigma$ result.  Statistically, the most significant sinusoidal
($n=1$) pulse was in the 1--20~keV dataset (340 photons), and had a
period $P=73.1$~ms with a $Z_1^2=34.4$ in 8365396 trials, a 34\%
chance of being a false positive.  Assuming that this signal is not
significant, we derive an upper limit on the pulse fraction of 45\%
for a sinusoidal pulse profile and 22.5\% for a $\delta$-function
pulse profile \citep{leahy83}.  Using the {\sc pn} count rate and size
of the Clump~2 and Diffuse regions, $\sim1/3$ of the counts in the
Clump~1 region is contamination from these regions.  Accounting for
this, we are only able to put an upper limit on the pulse fraction of
$67\%$ for a $\delta$-function pulse profile.  This upper limit is
consistent with the pulse fraction observed from other young neutron
stars.

\subsection {Australia Telescope Compact Array Observations}
\label{radio}

\begin{figure}
\begin{center}
\includegraphics[width=5.5in,angle=0]{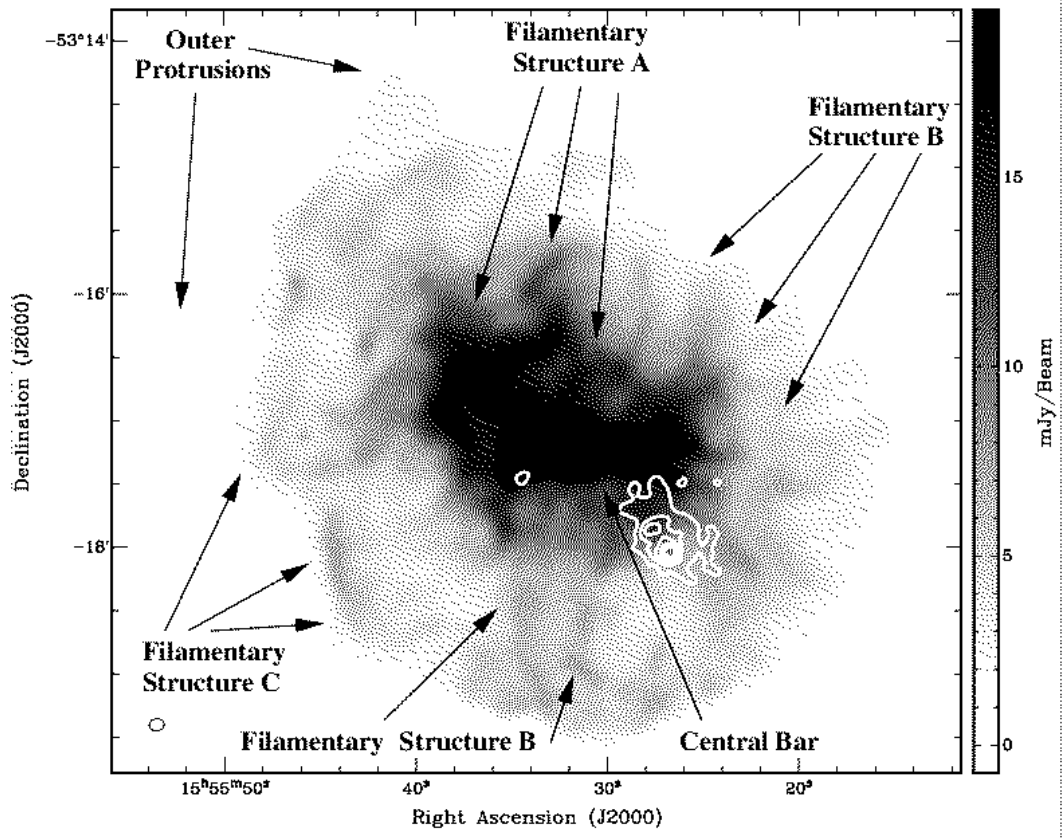}
\end{center}
\figcaption{1.4~GHz image of G328.4+0.2, overlaid with X-ray contours
  in green which represent 20\%, 35\%, ..., 90\% of the peak flux in
  the smooth X-ray image shown in Fig. \ref{xrayimg}.  The beam size
  of this image is 7\farcs0$\times$5\farcs8, and is shown in the lower
  left-hand corner of the image. The labels indicate examples of the
  different radio morphological features discussed in
  \S\ref{radio}. \label{radpic}}
\end{figure}

Based on previous radio observations of G328.4+0.2, there was a dispute
in the literature as to whether this source is a PWN, as argued by
\citet{gaensler00}, or a composite SNR, as argued by
\citet{johnston04}.  The argument for this source being a PWN centered
on the flat spectrum of the radio emission, as well as the high degree
of polarization observed from the center \citep{gaensler00}, while the
radial polarization angles observed at the edge of this is more
consistent with a SNR \citep{johnston04}.  If there is a SNR component
in G328.4+0.2, we expect that some of the radio emission from this
source should have a steep ($\alpha < -0.3$) spectrum.  To search for
such emission, we observed G328.4+0.2 for 12 hours at 1.4~GHz with the
Australia Telescope Compact Array (ATCA) on 2005~June~25.  Flux
density calibration was carried out using an observation of
PKS~B1934-638, and phase calibration was carried out with regular 
observations of PMN~J1603-4904.  The observation was carried out using
two 128~MHz bands, one centered at 1.344~GHz and the other at
1.432~GHz, and the data reduction was done using the {\sc miriad}
software package.  The observation was conducted when the ATCA was in
the 6B~configuration, which has a longest baseline of $\sim$6000~m
($\sim6\farcs9$) and a shortest baseline of $\sim$200~m
($\sim3\farcm4$). As a result, this dataset alone is not sensitive to
large-scale emission from G328.4+0.2.  To improve the sensitivity to
diffuse emission, we combined this dataset with the 1.4~GHz data used
by \citet{gaensler00} as well as continuum data gathered in the
Southern Galactic Plane Survey \citep{sgps}.  Total intensity images
from this combined dataset were formed using natural weighting,
multi-frequency synthesis, and maximum entropy deconvolution.   The
final image, shown in Fig.~\ref{radpic}, has a resolution of
7\farcs0$\times$5\farcs8, and an rms noise of
$\sim0.15$~mJy~beam$^{-1}$.  The measured 1.4~GHz flux density of
G328.4+0.2 is 13.8$\pm$0.4~Jy -- consistent with the value measured by
\citet{gaensler00}.  For a 4.5~GHz flux of $12.5\pm0.2$~Jy
\citep{gaensler00}, this implies that G328.4+0.2 has a radio spectral
index $\alpha=-0.03\pm0.03$.

As seen in Fig.~\ref{radpic}, the radio emission from G328.4+0.2 is 
very complicated, and contains multiple morphological features.  The
major features are:
\begin{itemize}
\item {\bf Central Bar:}  The Central Bar is the brightest
  morphological feature in G328.4+0.2, and has been previously
  detected at both 4.8~GHz \citep{gaensler00} and 19~GHz
  \citep{johnston04}.  The Central Bar runs roughly E-W, and the
  western edge of the bar is bifurcated, first noticed by
  \citet{johnston04}.  The length of the bar is $\sim1\farcm75$~long, and
  inside the bar there are three peaks in the radio emission.
\item {\bf Filamentary Structure~A:}  These are the curved filaments
  near the center of G328.4+0.2 which appear to be connected to the
  Central Bar, the brightest of which is the ``Y'' shaped structure NE
  of the eastern edge of the Central Bar.  In general, these filaments
  are more prominent on the eastern side of G328.4+0.2 and appear confined
  to the central region of G328.4+0.2, not extending much beyond the  
  inner half.
\item {\bf Filamentary Structure~B:}  These are the faint, radial
  filaments predominantly found on the western side of G328.4+0.2, as
  shown in Fig.~\ref{radfil}.  The inner parts of these filaments are
  located $\sim2\farcm5$ from the center of G328.4+0.2, and their length
  varies across G328.4+0.2 -- in the southern half, two filaments appear to
  extend to the edge of the source while in the northern and western
  parts of G328.4+0.2 they are substantially shorter.  While some of
  these filaments are kinked or curved, most are fairly straight and
  radial in orientation.  These features were not detected by
  \citet{gaensler00} and \citet{johnston04} due to the insufficient
  {\it u--v} coverage of those observations.
\item {\bf Filamentary Structure~C:}  As shown in Fig.~\ref{radpic},
  these are features located near the outer edge of G328.4+0.2 that
  are parallel to the outer edge.  These features are more prominent
  and more plentiful in the eastern half of G328.4+0.2.
\item {\bf Outer Protrusions:} The Outer Protrusions are faint features
  -- the two most prominent of which are in the NE quadrant of G328.4+0.2 -- 
  that extend beyond the outer boundary of G328.4+0.2.  Several of
  these structures have bow-shock morphologies.
\end{itemize}
The physical interpretation of these features will be presented in
\S\ref{obsres}.  It is worthwhile to note here that several of these
morphological features (e.g. the Central Bar and the internal filamentary
structures) have been observed in other PWNe such as MSH~15-5{\it 6}
\citep{dickel00} and 3C58 \citep{slane04}, while others (e.g. the
Filamentary Structure~C and Protrusions) are more characteristic of
SNRs, such as the Vela SNR \citep{bock98}.

This {\it XMM} observation also allows, for the first time, a comparison 
between the radio and X-ray morphology of G328.4+0.2.  As shown in 
Fig.~\ref{radpic}, there is a significant offset between the X-ray and the 
center of the radio emission, with Clump~1 located $\sim 80^{\prime
  \prime}$ from the center of the radio emission.  Additionally, the
extent of the X-ray emission is significantly smaller than that of the
radio emission.  A physical interpretation of the X-ray morphology and
its relation to the radio emission will be discussed in \S\ref{obsres}.   

\subsubsection{Spectral Index Map}
\label{specind}
The previous 1.4~GHz dataset had a resolution ($\sim20\arcsec$)
significantly worse than that of the 4.5~GHz data, and therefore is
not suitable for determining if there are small scale changes in
$\alpha$ inside G328.4+0.2.  With these new, high-resolution 1.4~GHz
observations, it was possible to make a spectral index map of
G328.4+0.2 using our new 1.4~GHz data and the 4.5~GHz data presented
by \citet{gaensler00} since these datasets have comparable {\it u--v}
coverage.  To detect any variation in $\alpha$, we made a spectral
tomography map of G328.4+0.2 \citep{katz-stone97}.  To do this, we
first produced a 1.4~GHz image of G328.4+0.2 from data matched in {\it
  u--v} coverage with the 4.5~GHz data, and then smoothed both the new
1.4~GHz image and the 4.5~GHz image to a resolution of 8\arcsec~to
account for the poorer resolution of the 1.4~GHz data. Finally, we
produced a series of difference images ($I_{\rm diff,\alpha}$) using
the following formula: 
\begin{eqnarray}
\label{diffimage}
I_{\rm diff,\alpha} & = & I_{\rm 1.4} - I_{\rm
  4.5}\left(\frac{1.4}{4.5}\right)^{\alpha} 
\end{eqnarray}
where $I_{\rm 1.4}$ and $I_{\rm 4.5}$ are the 1.4 and 4.5~GHz images
produced above.  In this method, the spectral index of a region is
determined by the spectral index at which it disappears from the
difference image.  As shown in Fig.~\ref{specimages}, most of the radio
emission from G328.4+0.2 has a spectral index between $\alpha \sim -0.1$ and
$\alpha \sim +0.1$, while the outer edges of G328.4+0.2 have a steeper
spectrum ($\alpha \sim -0.4$) than the center, particularly the
western edge of G328.4+0.2.  This steeper spectrum material is
coincident with some of the Filamentary Structure~C discussed in \S
\ref{radio}, but there are no spectral features associated with any of
the other radio morphological features in G328.4+0.2 or with the X-ray
emission.  A physical interpretation of these results will be
discussed in \S \ref{obsres}.

\subsection{Search for the radio pulsar at Parkes}
\label{pulsar}

As part of a project to search for pulsar counterparts to all Galactic
PWNe (e.g. \citeauthor{camilo02b}~\citeyear{camilo02b}), on 2005
October 13 we observed G328.4+0.2 using the ATNF Parkes telescope in
NSW, Australia.  As for similar such work, we employed the central
beam of the Parkes multibeam receiver at a central frequency of
1374\,MHz, with 96 frequency channels across a total bandwidth of
288\,MHz in each of two polarizations.  The integration time was
24\,ks, during which total-power samples were recorded every 0.25\,ms
for off-line analysis.

We analyzed the data with standard pulsar searching techniques using
PRESTO \citep{ransom02}.  We searched the dispersion measure range
0--2600\,cm$^{-3}$\,pc (twice the maximum Galactic DM predicted for
this line of sight by the \citeauthor{cordes02}~\citeyear{cordes02}
electron density model), while maintaining close to optimal time
resolution.  In our search we were sensitive to pulsars whose spin
period could have changed moderately during the observation due to
very large intrinsic spin-down.  The search followed very closely that
described in more detail in \citet{camilo06}.  We did not identify any
promising pulsar candidate in this search.

Applying the standard modification to the radiometer equation, for an
assumed pulsation duty cycle of 10\%, and accounting for a sky
temperature at this location of 15\,K, we were nominally sensitive to
long-period pulsars having a period-averaged flux density at 1.4\,GHz
of $S_{1400} >0.05$\,mJy.  In fact, this limit applies only to such
long-period pulsars as to not be of practical interest for us: for a
distance of $\sim 17$\,kpc along this line of sight, the expected DM
is $\approx 1200$\,cm$^{-3}$\,pc, and the scatter-broadening of the
radio pulses due to multipath propagation is expected to be $\sim
50$\,ms at 1.4\,GHz \citep{cordes02}.  This would likely render
pulsations undetectable from any short-period pulsar, such as we
expect to power G328.4+0.2, regardless of average radio flux.  We
therefore repeated the search at a higher radio frequency $\nu$, since
the scattering timescale is approximately $\propto \nu^{-4}$.

On 2007 January 4 we observed G328.4+0.2 at Parkes at a central frequency
of 3078\,MHz, with 288 channels spanning a bandwidth of 864\,MHz in
each of two polarizations.  The total-power samples were recorded every
1\,ms for 30\,ks, and analyzed in a manner analogous to that described
previously for the 1.4\,GHz data.  This time a few somewhat-promising
candidates were identified in the analysis, and a second 3\,GHz
observation was made, on 2007 March 19, for 36\,ks.  Analysis of
this second observation did not confirm the original candidates,
and we have therefore not detected any radio pulsar counterpart for
the PWN in G328.4+0.2.  The sensitivity of our 3\,GHz observations was
about 0.03\,mJy for long-period pulsars ($P \ga 20$\,ms) and decreasing
gradually for shorter periods.  For the predicted DM, at this frequency
the scattering timescale is expected to be $\sim 2$\,ms, comparable to
the dispersion smearing across each individual channel, $\sim 1$\,ms.
Propagation effects should therefore not have prevented the detection
of signals with $P \ga 5$\,ms.

Converting the 3\,GHz flux density limit to a frequency of 1.4\,GHz,
using a typical pulsar spectral index of --1.6 \citep{lorimer95},
results in $S_{1400} \la 0.1$\,mJy.  For a distance of $\sim 17$\,kpc,
this corresponds to a pseudo-luminosity limit of $L_{1400} \equiv
S_1400 d^2 \la 30$\,mJy\,kpc$^2$.  This is comparable to $L_{1400}$ of
the very young pulsars B1509--58, J1119--6127, and Crab
\citep{camilo02c}, but a factor of about 60 greater than for the young
pulsar in 3C58 \citep{camilo02a}, which has the smallest known
radio luminosity among young pulsars.  Based on these results, it is
therefore entirely possible that G328.4+0.2 harbors an as-yet
undetected young pulsar beaming toward the Earth with an ordinary
radio luminosity.

\section{Interpretation of X-ray and Radio Observations of G328.4+0.2}
\label{theory}

The X-ray spectrum of G328.4+0.2 is characteristic of PWNe (see
\citeauthor{gotthelf03} \citeyear{gotthelf03} for a compilation 
of the X-ray properties of PWNe), and therefore conclude that the
X-ray emission from G328.4+0.2 comes from a PWN and not from a SNR.
The same is true for the polarized, flat-spectrum radio emission
detected from the center of G328.4+0.2 which are also characteristic
of PWNe.  Therefore, in the following discussion we assume that the
X-ray and flat spectrum radio emission are both produced by a PWN.  In
\S\ref{evolution}, we discuss the evolutionary sequence of PWNe in
SNRs and the observational signatures of each stage.  In
\S\ref{obsres}, we use the results from the observations presented in
\S\ref{xray} and \S\ref{radio} to draw general conclusion about the
properties of G328.4+0.2. 

\subsection{Evolution of PWN in SNRs}
\label{evolution}

Both PWNe and SNRs are dynamic objects, and when the PWN is inside the
SNR its evolution is affected by the behavior of both the central
neutron star and the surrounding SNR.  While the PWN is inside the
SNR, it typically goes through three evolutionary phases
\citep{chevalier98,vdswaluw04}:  
\begin{itemize}
\item{\bf The Free-Expansion Phase -- }  In this phase, the PWN freely
  expands into the cold material inside the SNR, sweeping up and
  shocking the surrounding ejecta into a thin shell
  \citep{chevalier92,vdswaluw01a}.  Since the PWN is confined only by
  the shock wave its expansion drives into the surrounding SNR, and
  the velocity of this shock wave is much larger than the neutron star
  velocity, it is free to move inside the SNR with the neutron star.
\item {\bf Collision with the Reverse Shock -- } As the SN sweeps up
  and shocks the surrounding a ambient material, a reverse shock (RS)
  is driven into the ejecta.  Eventually, the PWN will encounter the
  RS, and as a result, can not continue to freely expand inside the
  SNR because it is no longer in an essentially pressureless
  environment.  Initially, the pressure behind the RS is higher  
  than the pressure inside the PWN, and as result the PWN is
  compressed.  As it contracts, the pressure inside the PWN increases
  adiabatically and eventually will be higher than its surrounding,
  and will as a result re-expand inside the SNR
  \citep{reynolds84,blondin01,vdswaluw01a,bucciantini03}.  Once the
  PWN encounters the RS, the expansion velocity of the PWN decreases 
  signficantly and falls below that of the neutron star, which is
  unaffected by this collision.  As a result, the neutron star can
  detach itself from its PWN.
\item {\bf Relic PWN Phase -- }  When the neutron star detaches from
  the relic nebula, it forms a new PWN from the relativistic
  $e^{+}/e^{-}$ plasma it continues to inject into the SNR
  \citep{vdswaluw04}.  The PWN around the neutron star and relic
  nebula evolve differently.  The relic nebula continues to
  contract/expand inside the SNR until it achieves pressure
  equilibrium with its surroundings, a process that can take many tens 
  of thousands of years.  The new PWN initially expands sub-sonically,
  but when the neutron star is $\sim2/3$ of the way to the SNR shell,
  its velocity will become supersonic relative to the surrounding
  material and the PWN will take on a bow-shock morphology
  \citep{vdswaluw04}.  Eventually, the neutron star will leave the
  SNR, and as it passes through the SNR shell it may re-energize the
  surrounding SNR material, as possibly observed in SNRs G5.4-1.2
  and CTB80 \citep{shull89}.
\end{itemize}

As the PWN evolves inside the SNR, its appearance changes radically.
During the Free-Expansion phase, the morphology of the PWN is
determined by the properties of the particle wind expelled by the
neutron star.  In these cases, the neutron star is in the center of
the PWN, and there is no significant offset between the radio and
X-ray emission from the PWN.  In general, during this stage the PWN is
located near the center of the observed SNR shell.  Examples of PWNe
in this evolutionary stage are the PWNe in SNR 0540-693 in the LMC
\citep{reynolds85}, as well as those in the Milky Way SNRs G11.2-0.3
\citep{tam02,roberts03} and G21.5-0.9 \citep{matheson05}. 

Due to the offset in the PWN's position with respect to the SNR's
center as a result of the neutron star's velocity or inhomogeneities
in the ISM, it is expected that one side of the PWN will encounter the
RS before the other side \citep{blondin01,vdswaluw04}.  As a 
result, the PWN will no longer be symmetrically oriented around the
neutron star \citep{vdswaluw04}, leading to an offset between the
radio and X-ray emission from the PWN.  Because the cooling time for
X-ray producing electrons is very short compared to that of radio
emitting electrons, the X-ray emission of a PWN is expected to be
brightest at the current location of the neutron star while the radio
emission reflects the effect of the RS on the PWN.  The
compression/re-expansion cycle triggered by the PWN/RS collision will
also affect the appearance of the PWN.  According to a spherically
symmetric MHD simulation of a PWN in this phase, compression of the
PWN by the RS leads to an over-pressurized region forming in the
center of the PWN.  Material injected by the neutron star after this
point is then confined to the small region, leading to the formation
of a radio/infrared ``hot spot'' in the center of the PWN
\citep{bucciantini03}.

The PWN/RS interaction also leads to the formation of hydrodynamic,
primarily Rayleigh-Taylor (R-T), instabilities at the PWN/SNR interface
\citep{blondin01}.  During the PWN's initial free-expansion phase, the
shell of material swept up by the PWN is subject to both thin shell
and R-T instabilities \citep{jun98,bucciantini04}, but the growth rate of
these features is expected to be sufficiently small that the PWN is not
disrupted, especially if even a small percentage of the total energy
of the pulsar's wind is in magnetic fields \citep{bucciantini04}.
However, during the PWN/RS interaction, rapid mixing of pulsar wind
and SNR material is expected when the PWN re-expands into the SNR
\citep{blondin01}.  In fact, numerical simulations suggest that the
PWN is disrupted after its first re-expansion into the SNR as a result
of these instabilities \citep{blondin01}.  It is important to note
that these instabilities are only expected to affect the relic nebula
and not the new PWN formed by the neutron star further from the SNR's
center.  Once the composite SNR enters the Relic PWN phase, the X-ray
emission is expected to be dominated by the new PWN since it contains
the high energy particles recently injected by the neutron star, while
the radio emission is dominated by the relic nebula which contains
most of the older particles that are expected to contribute at low
frequencies. As a result, during this phase the radio-emitting
electrons are expected to be dominated by electrons injected during
the free-expansion phase, while the X-ray is from electrons injected
after the passage of the RS.

\subsection{Observational Results}
\label{obsres}

In this Section, we use the results from the observations presented in
\S\ref{xray} and \S\ref{radio}, as well as the basic evolutionary
sequence for PWN in SNRs described above in \S \ref{evolution} to make
some initial statements on the nature, evolutionary state, and
properties of G328.4+0.2.  The discussion given below is a very general
interpretation of observed radio and X-ray features in G328.4+0.2,
it provides a framework in which to test the various scenarios for
G328.4+0.2 discussed in \S\ref{g328comp} and \S\ref{g328pwn}. 

\begin{figure}
\begin{center}
\includegraphics[width=5.0in,angle=-0]{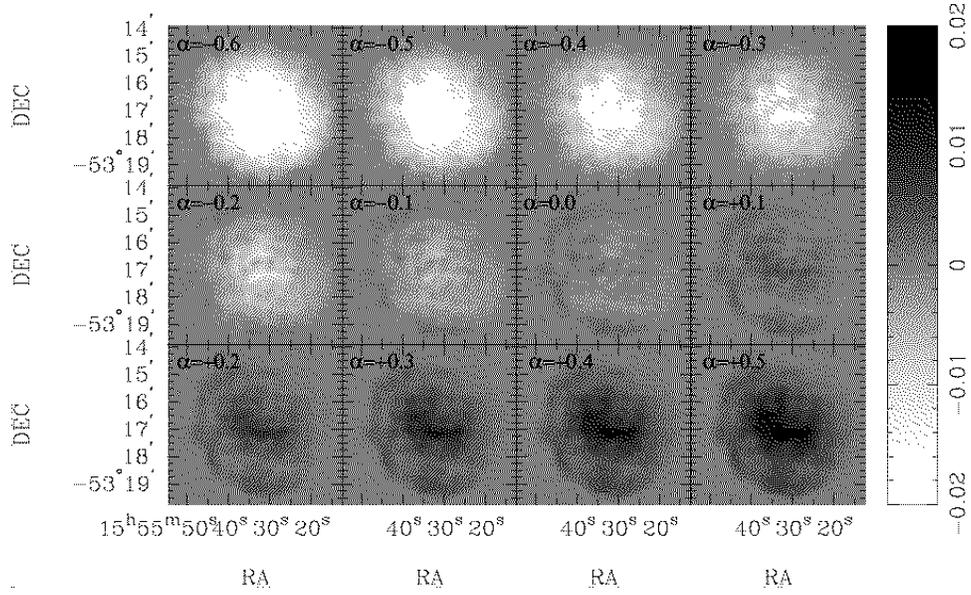}
\end{center}
\figcaption{Spectral tomography images of G328.4+0.2, as described in
  \S\ref{specind}.  The spectral index $\alpha$ is given in the upper
  left hand corner of each image, where $S_{\nu} \propto \nu^\alpha$.
  \label{specimages}}  
\end{figure}

As mentioned in \S \ref{intro}, there is a debate in the literature as
to whether G328.4+0.2 is a PWN \citep{gaensler00,hughes00} or a 
composite SNR \citep{johnston04}.  In neither of the X-ray or radio
observations presented above is there clear evidence (e.g. thermal
X-ray emission or a bright, steep spectrum, radio shell) for a SNR
component.  If G328.4+0.2 is a composite SNR, then the outer boundary
of the radio emission likely marks the outer radius of the SNR
component, while the observed flat-spectrum radio emission and power
law X-ray emission are emitted by the PWN.  Using the extent of the
flat-spectrum radio emission shown in Fig.~\ref{specimages} to
estimate the size of the PWN component, we obtain that the radius of
the PWN in this source must be $\ga 2/3 R_{\rm G328}$, the radius of
G328.4+0.2.  If G328.4+0.2 is a composite SNR, then Filamentary
Structure 3, which as mentioned in \S\ref{specind} might have a
steeper spectral index than the rest of the radio emission in
G328.4+0.2, would be emission from the SNR.  This emission is then
possibly analogous to the corrugated structures seen in the NE part of
the Tycho's SNR \citep{velazquez98}.  Additionally, in this case the
Outer Protrusions mentioned in \S\ref{radio} maybe are ejecta
``bullets'', similar to those observed in the Vela SNR
\citep{aschenbach95} and SNR~N63A \citep{warren03}. 

If G328.4+0.2 is a PWN, then the outer boundary of the radio emission
is the outer radius of the PWN and the SNR in which it resides is
undetected, similar to the case for the Crab Nebula.  As a result, the
steep spectrum radio emission seen at the edge of G328.4+0.2, as well
as Filamentary Structure C and the Outer Protrusions observed in the
radio, correspond to material swept-up by the PWN.  This is because
radio emission from the pulsar wind is observed to have a flat
spectrum, unlike this material.  As a result, Filamentary Structure C,
as well as the radial component seen by  \citet{johnston04} in the
polarization angle along the outer edge of G328.4+0.2, are the result
of the hydrodynamical (HD) instabilities at the PWN/SNR interface.
Since these instabilities only occur when the PWN is accelerating the
shell of swept-up material surrounding it, he current pressure
inside the PWN $(P_{\rm pwn})$ must be higher than that of the 
SNR material just of the PWN $[P_{\rm snr}(R_{\rm pwn})]$.
Additionally, in this case the Outer Protrusions would be the result of
the PWN currently expanding into a clumpy medium (e.g. the PWN analog
of the process described for young SNR by \citeauthor{jun96}
\citeyear{jun96}), which requires that the expansion speed of the PWN,
$v_{\rm pwn}$, is currently positive.  In this case, it is possible
that the SNR surrounding G328.4+0.2 will be detected at a later date,
as was the case for G21.5-0.9 \citep{matheson05}.

Regardless of whether G328.4+0.2 is a composite SNR or a PWN, the offset
between the radio and X-ray emission from the PWN component implies
that the PWN/RS collision has already occurred.  The Central Bar is
then the remains of the over-pressurized region created when the PWN
was compressed by the RS, with the bar-like shape of this region the
result of either an asymmetric RS or the anisotropic wind emitted by
the neutron star.  The Central Bar should then consist of pulsar wind
material, which accounts for the $\alpha \sim 0$ spectral index of
this region as shown in Fig.~\ref{specimages}.  Additionally, the flat
spectral index observed from Filamentary Structures A implies that
this feature is emitted by pulsar wind material.  As mentioned in
\S\ref{evolution}, when the PWN re-expands after the initial
compression by the RS, numerical simulations suggest that the rapid
mixing of the SNR ejecta and pulsar wind material can then occur.  In
their 2D simulations of this process, \citet{blondin01} observe
features similar to that of Filamentary Structure A, and as a result
we conclude that the existence of these features requires that the PWN
has re-expanded at least once after its initial compression by the RS.
Since the radius of flat-spectrum radio emission from G328 is larger
than the outer radius of Filamentary Structure A, the instabilities
formed during this expansion must not have completely disrupted the
PWN.  This differs from the results of \citet{blondin01}, in which the
PWN is disrupted during the first re-expansion.  This is most likely
due to the damping effect of the PWN's magnetic field on
Raleigh-Taylor instabilities, which is not accounted for by
\citet{blondin01}. 

Finally, we discuss the X-ray emission seen from G328.4+0.2 which,
based on its X-ray spectrum, is emitted by pulsar wind material.  The
observed offset between Clump~1, Clump~2, and the Diffuse regions of
the X-ray emission implies that the PWN is not freely expanding,
consistent with the explanation that the PWN has collided with the RS.
We identify Clump~1 as the current location of the neutron star since
it is the brightest X-ray feature, Clump~2 as the location of the
termination shock in the PWN \citep{kennel84}, and the Diffuse
emission is produced by recently injected plasma streaming away from
the neutron star.  In this case, the extent of the Diffuse component
depends on the synchrotron lifetime of the X-ray emitting particles in
the PWN.

The X-ray emission also provides an estimate of the physical properties
of the central neutron star, namely a measure of the neutron star's
rotational spin-down energy, $\dot{E}$.  A comparison of observed
X-ray luminosity $L_x$ and $\dot{E}$ shows a trend that neutron stars
with a higher $L_{x}$ have a higher $\dot{E}$, and that the
relationship between these two quantities is \citep{possenti02}:
\begin{eqnarray}
\label{lxedot}
\log L_{\rm X,(2-10)} & = & 1.34 \log \dot{E} - 15.34
\end{eqnarray}
where $L_{\rm X,(2-10)}$ is the X-ray luminosity of the source between 
2 and 10~keV, albeit with a significant scatter \citep{possenti02}.
For the absorbed power-law fit to the X-ray emission from G328.4+0.2
given in Table \ref{specres}, the unabsorbed 2--10~keV flux of
G328.4+0.2 is $\sim1\times10^{-12}$~ergs~cm$^{-2}$~s$^{-1}$.  For a 
distance to G328.4+0.2 of $d=17 d_{17}~{\rm kpc}$, we obtain that:
\begin{eqnarray} 
\label{luminxray}
L_{\rm X,(2-10)} & \sim & 3.5d_{17}^2 \times 10^{34}~{\rm ergs~s}^{-1},
\end{eqnarray}
which, using Eq. (\ref{lxedot}), gives us an estimate of $\dot{E}$:
\begin{eqnarray}
\label{edot}
\dot{E} \sim 1.7d_{17}^{1.49}\times10^{37}~{\rm ergs~s}^{-1}.
\end{eqnarray}
This number is somewhat less than estimate obtained
by\citet{gaensler00} ($\dot{E}=8.3\times10^{38}~{\rm ergs}~{\rm
  s}^{-1}$), who assumed that $\dot{E} = 4\times10^{-4} L_{R}$, where
$L_R$ is the radio luminosity of G328.4+0.2.  Our estimate of
$\dot{E}$ is similar to the estimate by \citet{hughes00}
($\dot{E}\sim10^{37}-2\times10^{38}~{\rm ergs}~{\rm s}^{-1}$), who
also used the $L_x-\dot{E}$ relation.    

\section{Simple Hydrodynamic Model for the Evolution of a PWN
  inside a SNR} 
\label{model}

In order to determine what neutron star, SN, and ambient density
properties are required to produce a system with these properties
described in \S\ref{obsres}, we have developed a simple hydrodynamic
(HD) model for the evolution of a PWN inside of a SNR, which we then
apply to G328.4+0.2 in \S\ref{modelapp}.  This model is based largely
on the models developed by \citet{blondin01} and \citet{vdswaluw01a}.
The main goal of this model is to determine the radius of the PWN,
$R_{\rm pwn}$, as it progresses through the evolutionary sequence
described in \S\ref{evolution}.  In this model, we assume that the PWN
can be treated as a perfect gas with adiabatic index $\gamma=4/3$ and
that is expanding into a SNR filled with a perfect gas with adiabatic
index $\gamma=5/3$.  We also assume that the material swept-up by the
PWN initially lies in a thin shell with inner radius $R=23/24~R_{\rm
  pwn}$ \citep{vdswaluw01a}, as shown in Fig.~\ref{snrpwnstruct}.  The
dynamics of this mass shell are determined by the difference in
pressure between the PWN and SNR, and by calculating the radius of
this mass shell we determine $R_{\rm  pwn}(t)$.  Once the PWN enters
the Relic PWN phase of its evolution, this model only determines the
properties of the relic nebula.  What follows is a brief qualitative 
description of the model used, while the full suite of equations used
to implement it quantitatively can be found in Appendix \ref{model2}.

As mentioned above, we model $R_{\rm pwn}(t)$ by calculating the outer
radius of the mass shell swept up by the PWN, ignoring the effect of
any instabilities which could disrupt this shell.  We solve for
$R_{\rm pwn}(t)$ by assuming that we know the values for the relevant
quantities at a time $t-\Delta t$, and then calculate them for a time
$t$, since the relevant equations can not be solved at all times
analytically.   we wrote a program in {\sc IDL} to implement this
numerically using the following procedure: 
\begin{enumerate}
\item Calculate $R_{\rm pwn}(t+\Delta t)$ by assuming that the mass
  shell around the PWN between $t$ and $t+\Delta t$ moves with a
  constant velocity $v_{\rm pwn}(t)$.
\item Calculate the internal energy of the PWN, $E_{\rm pwn}(t+\Delta
  t)$, using the first law of thermodynamics:
  \begin{eqnarray}
    \label{epwneqn}
    \Delta E_{\rm pwn} & = & \dot{E}t - P_{\rm pwn}\Delta V_{\rm pwn}
  \end{eqnarray}
  which takes into account energy losses from the adiabatic
  expansion/contraction of the PWN as well as any energy input from
  neutron star into the PWN between $t$ and $t+\Delta t$ if the
  neutron star is still inside the PWN.   Since we assume the PWN is
  filled with a $\gamma=4/3$ perfect gas, $E_{\rm pwn} \propto P_{\rm
  pwn} V_{\rm pwn}$ from the ideal gas law, where $P_{\rm pwn}$ is the
  internal pressure of the PWN and $V_{\rm pwn}$ is the volume of the
  PWN, as defined in Equations (\ref{p_pwn} and (\ref{v_pwn}).  Since
  $V_{\rm pwn} \propto R^3$, and $\gamma = 4/3$ requires that $P_{\rm
  pwn} \propto V_{\rm pwn}^{-4/3}$, we derive that $P_{\rm pwn}
  \propto R^{-4}$.  As a result, if there is no input from the neutron
  star, then $E_{\rm pwn} \propto R_{\rm pwn}^{-1}$.  The energy input
  from the neutron star ($\Delta E_{\rm psr}$) is calculated by
  integrating Eq. (\ref{edoteqn}) between $t$ and $t+\Delta t$. 
\item Calculate $P_{\rm pwn}(t+\Delta t)$ using Equations
  (\ref{p_pwn}) and (\ref{v_pwn}). 
\item Calculate the pressure inside the SNR ($P_{\rm snr}$), the
  density inside the SNR ($\rho_{\rm ej}$), the velocity of the
  material inside the SNR ($v_{\rm ej}$), and the sound speed of the
  material inside the SNR ($c_s$), at the outer radius of the PWN,
  $R=R_{\rm pwn}(t+\Delta t)$ using a model for the evolution and
  structure of a SNR, as described in Appendix \ref{model2}.
\item If the PWN is expanding faster than the SNR material around it,
  increase the mass of the shell surrounding the PWN, $M_{\rm
  sw,pwn}(t+\Delta t)$, accordingly, as described in Appendix
  \ref{model2}.
\item Calculate the force on the mass shell surrounding the PWN,
  $F_{\rm pwn}(t+\Delta t)$, using Eqs. (\ref{fdelp}) and
  (\ref{forpwn}).  During the initial free-expansion of the PWN inside
  the SNR, these equations reduce to Equation A4 of
  \citet{vdswaluw01a} and Equation 14 of \citet{chevalier05}. 
\item Calculate the new velocity of the mass shell around the PWN,
  $v_{\rm pwn}(t+\Delta t)$, assuming that any mass swept up by the
  PWN between $t$ and $t+\Delta t$ is done so inelastically:
  \begin{eqnarray}
    \label{vpwneqn}
    v_{\rm pwn}(t+\Delta t) & = & \frac{M_{\rm sw}(t)v_{\rm pwn}(t) +
    F_{\rm pwn}(t+\Delta t) \Delta t }{M_{\rm sw}(t+\Delta t)} 
  \end{eqnarray}
  This is believed to be a reasonable approximation because the newly
  swept-up material is shocked by the mass shell and, as a result, its 
  pre-existing momentum is transferred to the internal energy of the
  mass shell.
\end{enumerate}

\begin{figure}
\begin{center}
\includegraphics[width=3.75in,angle=90]{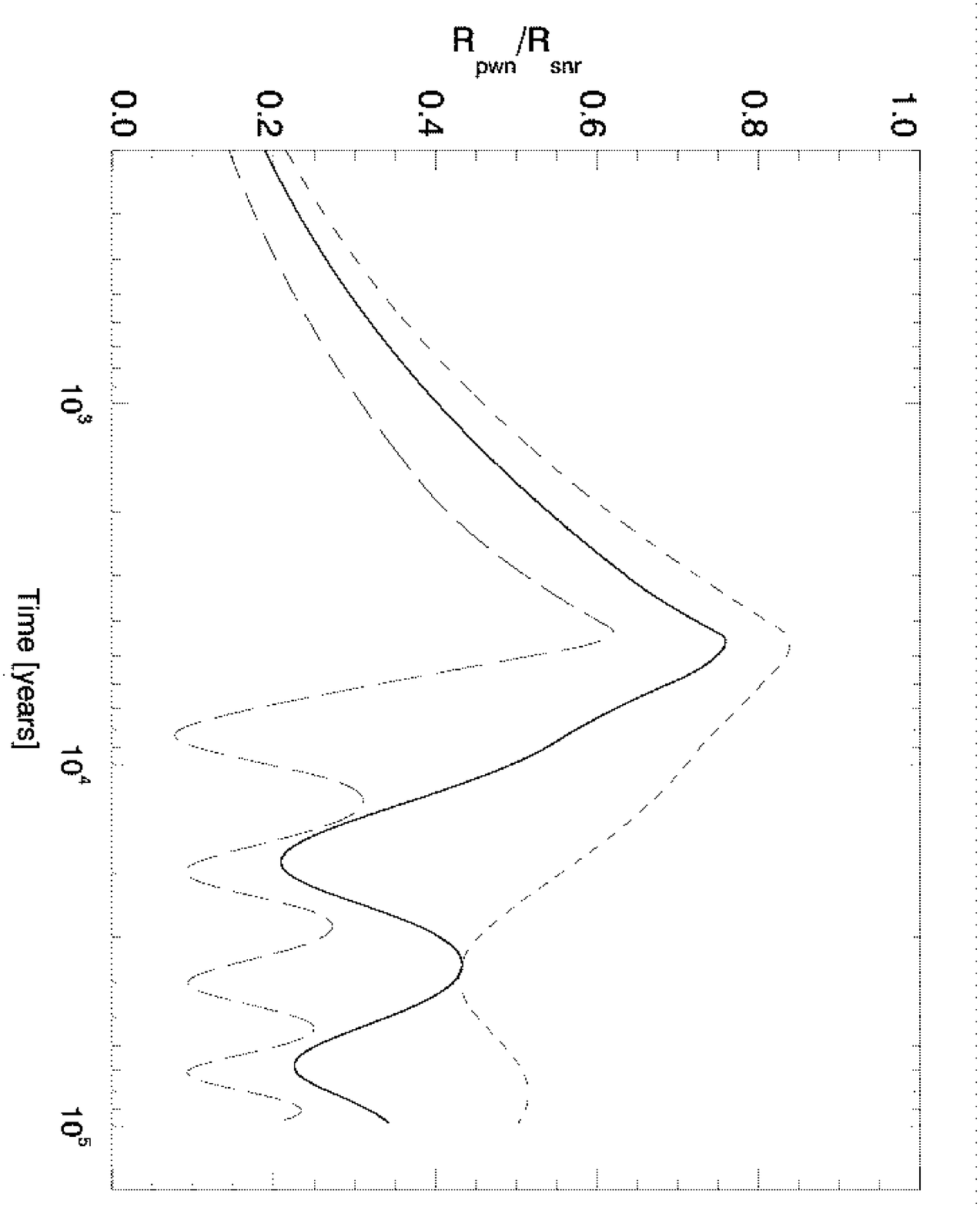}
\includegraphics[width=5.00in,angle=0]{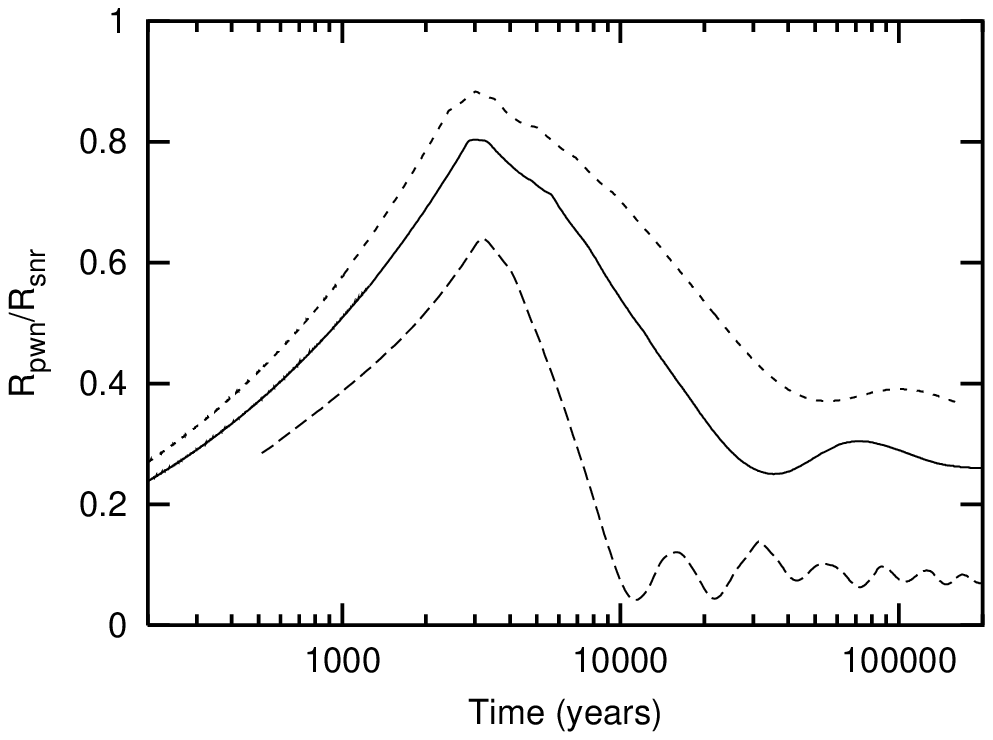}
\end{center}
\figcaption{$R_{\rm pwn}/R_{\rm snr}$ for Models A ({\it solid}), B
  ({\it long-dashed line}), \& C ({\it short-dashed line}) in
  \citet{blondin01}.  The {\it top} plot shows the result of the model
  presented in \S\ref{model}, while the bottom is a reproduction of
  Fig.~3 by \citet{blondin01}, reproduced by permission of the
  AAS. \label{blondinres}}
\end{figure}

This model is assumes that both the SNR and PWN are spherically
symmetric, the PWN remains centered on the center of the SNR at all
times, the PWN has no effect on the evolution of the SNR, and that the
material swept-up by the PWN is incompressible and has a negligible
internal pressure.  Additionally, this model ignores the effects of
magnetic field \citep{bucciantini03} and RT instabilities at the
PWN/SNR interface \citep{blondin01,vdswaluw04} on the properties of
the PWN.  Despite these simplifications, our model does a reasonably
good job of reproducing the results for Model~A in \citet{blondin01},
as shown in Fig.~\ref{blondinres}.  In general, relative to results of
\citet{blondin01} and other authors, our model tends to result in
larger oscillations in $R_{\rm pwn}/R_{\rm snr}$ and a larger
initial compression.  The first discrepancy results from neglecting
the effect of instabilities at the PWN/SNR interface that damp these
oscillations, and the second from not including the effect of
reflected shocks that enter the PWN at the time of the PWN/RS
collision \citep{blondin01}.  Additionally, we find that scenarios
with the same total amount of energy deposited by the neutron star
into the PWN, $E_{\rm psr}$ but with different neutron star properties
(e.g. different values of $P_0$ and $B_{\rm ns}$) produce the same
behavior of $R_{\rm pwn}(t)$.  This is different than the conclusion
of \citet{blondin01}, and believe that this discrepancy is the result
of using a more realistic expression for $\dot{E}$,
Eq. (\ref{edoteqn}), than a step function, the form used by
\citet{blondin01}.

\subsection{Application of Model to G328.4+0.2}
\label{modelapp}

In the following discussion, we use the model given in \S\ref{model}
for a PWN's evolution inside a SNR to examine the different
possibilities for the nature of G328.4+0.2 given in \S\ref{obsres}.
We first analyze the possibility that G328.4+0.2 is a composite SNR
(\S\ref{g328comp}), and then he possibility that G328.4+0.2 is a
PWN (\S\ref{g328pwn}).  This model requires six inputs: the
characteristic timescale of the neutron star's spin-down, $\tau_0$,
initial spin-down power of the neutron star $\dot{E_0}$, the velocity
of the neutron star, $v_{\rm ns}$, the kinetic energy of the SN ejecta
$E_{\rm sn}$, the mass of the SN ejecta $M_{\rm ej}$, and the number
density of the surrounding material $n$.  In order to calculate these
values, we use the following information:
\begin{itemize}
\item The distance to G328.4+0.2 is 17~kpc ($d_{17} \equiv 1$), which
  is the lower limit on the distance to this source as determined by
  \citet{gaensler00} using H{\sc i} absorption.  This implies that the
  current radius of G328.4+0.2 is $R_{\rm G328}\equiv12.5~{\rm pc}$.
\item The neutron star inside G328.4+0.2 is spinning down with a braking
 index $p=3$, the braking index produced by a pure dipole surface
 magnetic field.  Additionally, we assume that the neutron star's
 moment of inertia is $I=10^{45}~{\rm g~cm}^2$, the value derived for
 standard equations of state for a neutron star \citep{shapiro83}.
 Both of assumptions are standard in the literature
 (e.g. \citeauthor{blondin01} \citeyear{blondin01}).  
\item The offset between the Clump~1, which as described in
  \S\ref{obsres} is believed to be the location of the neutron star
  powering the PWN, and the center of the radio emission is due to
  neutron star's spatial velocity, $v_{\rm ns}$.  Using \S\ref{image},
  we determine that this observed offset corresponds to physical
  distance of $\sim6.6d_{17}$~pc.  Since the observed offset is due only
  to the neutron star's velocity in the plane of the sky, it is a
  lower limit on the true distance the neutron star has traveled since
  the SN explosion, $r_{\rm ns}$.  If we assume that $v_{\rm
  ns}=r_{\rm ns}/t_{\rm now}$, where $t_{\rm now}$ is the age of
  G328.4+0.2, the observed offset allows us to estimate the minimum
  spatial velocity of the neutron star, $v_{\rm ns}^{\rm min}$, equal to:
  \begin{eqnarray}
    \label{vnsmin}
    v_{\rm ns}^{\rm min} & = & \frac{6.6d_{17}~{\rm pc}}{t_{\rm now}}.
  \end{eqnarray}
\item Using Equation (\ref{tauphysical}), we determine that for
  standard initial periods ($P_0 \sim 5 - 20$~ms) and magnetic field
  strengths ($B_{\rm ns} = 5\times10^{11} - 10^{13}$~G), $\tau_0$
  varies from $\sim100-2000$ years.  To cover this range, we assume
  that $\tau_0$ of the neutron star in G328.4+0.2 can have one of
  three different values: 
  \begin{eqnarray}
    \label{tauvals}
    \tau_0 = 430, 770,~{\rm and}~1730~{\rm years}.
  \end{eqnarray}
  which respectively correspond to a neutron star with $B=10^{12}$~G
  and $P_0=5$~ms, $B=3\times10^{12}$~G and $P_0=20$~ms, or
  $B=5\times10^{11}$~G and $P_0=5$~ms.  This range of $\tau_0$ is
  similar to those used by \citet{blondin01}, \citet{vdswaluw01a}, and
  \citet{bucciantini03}.
\item G328.4+0.2 is expanding into a uniform medium ($s=0$ in the
  notation of \citeauthor{chevalier82} \citeyear{chevalier82}).  This 
  assumes G328.4+0.2 is much larger than either the main sequence and
  late-stage wind bubble formed by its progenitor, both of which are
  expected to have an interior $\rho \propto r^{-2}$ density structure.  
  While the typical size of these structures is smaller than $R_{\rm
  G328}$, this is not the the case for very massive stars ($M \ga
  15~M_{\odot}$) for which these bubbles can reach sizes of
  $\sim100$~pc or larger in low density ($n \la 1~{\rm cm}^{-3}$)
  environments \citep{chevalier89,chevalier89b}.  In this case, our
  assumption of a constant density medium would not be correct.
  However, the effect of G328.4+0.2 still being inside a stellar wind
  bubble since this does not significantly modify the evolution of the
  SNR.  Since their no {\it a priori} information on the density around
  G328.4+0.2, we assume it is one of the assume it has one of the
  following values:  
  \begin{eqnarray}
    \label{nvals}
    \log n = -1.5,-1.0,0,0.5~{\rm cm}^{-3}.
  \end{eqnarray}
  which cover the range of densities in the warm ionized medium and
  the warm neutral medium.
\item The ejecta mass of the SN explosion that formed G328.4+0.2, $M_{\rm
  ej}$ has one of the following values:
  \begin{eqnarray}
    \label{mejvals}
    M_{\rm ej} = 1, 5, 10 M_{\odot}
  \end{eqnarray}
  and that the kinetic energy of the ejecta, $E_{\rm sn}$ is:
  \begin{eqnarray}
    \label{esnvals}
    \log (E_{\rm sn}/10^{51}{\rm ergs}) = -0.5, 0, 0.5.
  \end{eqnarray}
  This range of $M_{\rm ej}$ and $E_{\rm sn}$ incorporate the range
  inferred from observations of ``normal'' SNe, but do not include
  hypernovae.
\item We assume that the $L_x-\dot{E}$ relationship used in
  \S\ref{obsres} is accurate to better than two orders of magnitude.
  As a result, in \S\ref{g328comp}, we assume that the current
  spin-down luminosity of the neutron star in G328.4+0.2 is one of the
  following: 
  \begin{eqnarray}
    \dot{E} & = & 0.017, 0.17, 1.7, 17,~{\rm
    and}~170\times10^{37}~{\rm ergs~s}^{-1},
  \end{eqnarray}
  and in \S\ref{g328pwn} assume that $1.7\times10^{35} < \dot{E} <
  1.7\times10^{39}$~ergs~s$^{-1}$. 
\end{itemize}
These are the initial conditions used in both \S\ref{g328comp} and
\S\ref{g328pwn}.  The remaining input parameters into the model are the
initial spin-down luminosity $\dot{E_0}$ and space velocity $v_{\rm
  ns}$ of the neutron star, and the method for determining the
possible values of these parameters is given in \S \ref{g328comp} and
\S\ref{g328pwn}. Finally, for all trials discussed in \S\ref{g328comp}
and \S\ref{g328pwn} the model begins at a time $t=0.5$~years, with
$\Delta t=0.5$~years.

\subsubsection{G328.4+0.2 as a Composite SNR}
\label{g328comp}
In this Section, we evaluate the possibility that G328.4+0.2 is a
composite SNR.  To do this, we first assume that the outer edge of the
radio emission from this source denotes the edge of the SNR, and
therefore $R_{\rm G328} \equiv R_{\rm snr}$.  As a result, for a given
value of $E_{\rm sn}$, $M_{\rm ej}$, and $n$, we use the model for the
evolution of a SNR discussed in Appendix~\ref{model2} to calculate the
current age of G328.4+0.2, $t_{\rm now}$.  With this value of $t_{\rm
  now}$ and assumed values for the current spin-down luminosity of the
neutron star, $\dot{E}$, and $\tau_0$, we are able to calculate both
the initial spin-down luminosity $\dot{E_0}$ and initial period $P_0$
of the neutron star in G328.4+0.2 using Eqs.~(\ref{edoteqn}) and
(\ref{e0dot}), respectively.

Using this procedure, we ran our model using all possible combinations
of the input parameters ($\tau_0$, $\dot{E}$, $E_{\rm sn}$, $M_{\rm
  ej}$, and $n$) given in \S\ref{modelapp}, for a total of 540
different combinations.   To see which combination of these input
parameters provide a plausible explanation for G328.4+0.2, we require
the following:
\begin{itemize}
\item {\bf Criterion 1:} $v_{\rm ns}^{\rm min} < 2000~{\rm
  km~s}^{-1}$, since a neutron star with a higher velocity than this is
  extremely implausible based on pulsar observations \citep{hobbs05}.
\item {\bf Criterion 2:} $P_0 > 2~{\rm ms}$, the minimum rotation
  period of a young proto-neutron star before it breaks up \citep{goussard98}.
\item {\bf Criterion 3:} The PWN is smaller than the SNR for all $t <
  t_{\rm now}$.  While it is possible that PWN could expand to fill
  the entire SNR, it is not considered likely, and is contrary to the
  model assumption that the PWN does not affect the evolution of the
  SNR.  Additionally, we also require that  $R_{\rm pwn}(t_{\rm now})
  \geq 0.67 R_{\rm snr}(t_{\rm now})$, due to the large observed size
  of the flat-spectral index radio emission which is produced from the
  PWN, as described in \S\ref{obsres}.
\item {\bf Criterion 4:} G328.4+0.2 is currently in the Free-Expansion 
  or Sedov-Taylor phase of its expansion, i.e.\ $t_{\rm now} < t_{\rm
  rad}$, where $t_{\rm rad}$, the age when a SNR goes radiative, is
  defined in Eq. \ref{trad}.  Once a SNR has entered its Radiative
  phase, it is expected that the radio emission from the SNR be
  confined to thin, bright, filaments like those observed in SNR G6.4-0.1
  \citep{mavromatakis04} -- which are not observed in G328.4+0.2, or
  that the SNR is radio-quiet.
\item {\bf Criterion 5:} The PWN/RS collision has already occurred, as
  described in \S\ref{obsres}, and the PWN has been compressed as
  a result of its collision with the RS.  This is required to explain
  the Central Bar, as described in \S\ref{obsres}.  This requires that
  $v_{\rm pwn}<0$ at some point in the past -- which can only occur at
  a time $t>t_{\rm col}$, the time when the PWN and RS collide.
\item {\bf Criterion 6:} The Central Bar created by the compression is
  still observable.  This is satisfied if either the PWN is currently
  being compressed, $v_{\rm pwn}(t_{\rm now}) < 0$, or if the
  compression ended sufficiently recently such that it can be
  observed.  The Central Bar is believed to be formed by both a
  pressure and a magnetic field enhancement at the center of the PWN
  \citep{bucciantini03}.  As a result its observable lifetime is the 
  synchrotron lifetime of electrons accelerated by the magnetic field
  enhancement.  Therefore, we assume that the lifetime of the central
  bar is the synchrotron age of the accelerated electrons, $\tau_{\rm
  synch}$, equal to:
  \begin{equation}
    \label{tsynch}
    \tau_{\rm synch} = 3\times10^{4} \nu^{-1/2} B_{\rm
    pwn}^{-3/2}~{\rm years}, 
  \end{equation}
  where $\nu$ is the observed frequency, in units of Hz, and $B_{\rm
  pwn}$ is the strength of the magnetic field inside of the PWN, in
  units of G.  With this information, in the case that $v_{\rm
  pwn}(t_{\rm now}) > 0$ we determine if the central bar is still
  observable by evaluating $\tau_{\rm synch}$ at 22~GHz (since this is
  the highest frequency at which the Central Bar is observed;
  \citeauthor{johnston04}~\citeyear{johnston04}) at the time when the 
  compression ends, assuming that $B_{\rm pwn}$ can be derived using
  the minimum energy estimate.   This criterion is satisfied if
  $\tau_{\rm synch} > t_{\rm now} - t_{\rm re-exp}$, where $t_{\rm
  re-exp}$ is the time when the compression phase ends.
%\item {\bf Criterion 7:}  The PWN/RS must have been unstable to R-T
%  instabilities sometime after the PWN/RS collision in order to
%  explain the Filamentary Structure~A and Internal Radio Filaments, as
%  discussed in \S\ref{obsres}.  This is implemented by requiring
%  that, for some $t>t_{\rm col}$, the pressure inside the PWN is
%  higher than that of the surrounding SNR, $P_{\rm pwn} > P_{\rm
%  snr}(R_{\rm pwn})$.
\end{itemize}
That the above criteria fall into two categories: criteria required
for physical plausibility (Criteria 1--3) and those which depend on
our interpretation of the radio and X-ray properties of G328.4+0.2
(Criteria 4--6). 

\begin{figure}
\begin{center}
\includegraphics[width=4.0in,angle=90]{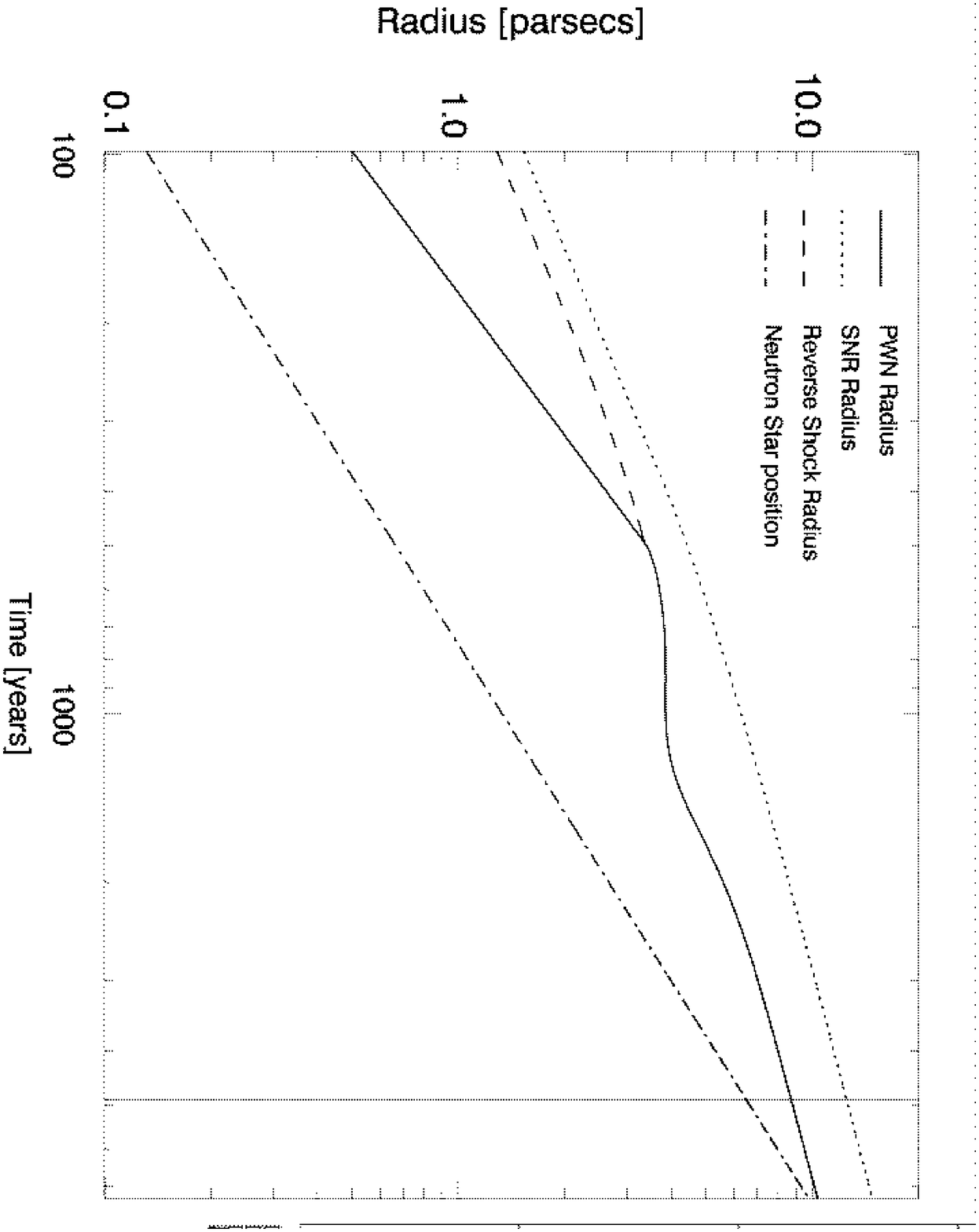}
\end{center}
\figcaption{The radius of the PWN, SNR, and SNR reverse shock as well
  as the location of the neutron star as a function of time if
  G328.4+0.2 is a composite SNR.  The vertical line indicates the
  current age of the system, and the properties of this system are
  given in Table \ref{modg328comp}. \label{figg328comp}} 
\end{figure}

Of the 540 possible combinations of the input parameters, only one
passes all six criteria.  The predicted SNR, PWN, and neutron star
properties of this scenario are given in Table \ref{modg328comp}, and 
the behavior of $R_{\rm pwn}$ as a function of time is given in Figure
\ref{figg328comp}.  In this scenario, G328.4+0.2 is quite young,
$\sim4900$~years old, and the energy injected into the SNR by the
neutron star is similar to the kinetic energy of the SN explosion
($\sim10^{51}$~ergs).  However, in this scenario, as shown in
Fig. \ref{figg328comp}, the predicted compression is very small;  when
the compression begins, $R_{\rm pwn}=3.838$~pc, and when the PWN
begins to re-expand into the SNR, $R_{\rm pwn}=3.834$~pc.  This small
decrease is not surprising given that, as shown in Table
\ref{modg328comp}, the total energy inputed into the PWN by the pulsar
($E_{\rm psr}$) is very close to the kinetic energy of the SN ejecta
($E_{\rm sn}$).  This negligible decrease in the volume of the PWN is
unlikely to form a central bar as prominent as the one observed
(Fig.~\ref{radpic}), and therefore we feel is unlikely to be the
correct explanation for G328.4+0.2.

\subsubsection{G328.4+0.2 as a PWN}
\label{g328pwn}

In order to evaluate if G328.4+0.2 is a PWN, we use the model presented in
\S\ref{model} to determine the earliest time\footnote{Due to
  oscillations in radius the PWN undergoes after its collision with
  the RS, it can reach the current size at multiple times.}
($t_{\rm now}$) at which a PWN powered by a neutron star with a given
initial period $P_0$ reaches the observed size of G328.4+0.2 ($R_{\rm
  pwn}=12.5$~pc) if it is expanding into as yet unseen SNR formed by
ejecta with initial mass $M_{\rm ej}$ and kinetic energy $E_{\rm sn}$
which exploded in a constant-density ambient medium with number
density $n$.  To consider all reasonable cases, we ran our model using
all combinations of the values of $\tau_0$, $E_{\rm sn}$, $M_{\rm ej}$,
and $n$ given in \S\ref{modelapp}, as well as $P_0=5, 10, 25, 100$~ms
for a total of 432 different trials.  In this scenario, since it is
not possible to determine an independent estimate of the age of the
system, it is necessary to assume a value of $P_0$.  To determine
which of these combinations are possible explanations for G328.4+0.2,
we required that: 
\begin{itemize}
\item {\bf Criterion~1}: $v_{\rm ns}^{\rm min}<2000~{\rm km}~{\rm
  s}^{-1}$.  Since, in this scenario, we have no prior estimate of the
  age of G328.4+0.2, when we run our model we assume that $v_{\rm ns}
  = 0$.   This does not effect our results because $r_{\rm ns}$, as
  measured in \S\ref{modelapp}, is less than $R_{\rm pwn} \equiv
  R_{\rm G328}$, and therefore the neutron star is always injecting
  energy into the PWN as it does if $v_{\rm ns} = 0$.  Once, for a
  given set of input parameters we have determined $t_{\rm now}$, we
  calculate $v_{\rm ns}^{\rm min}$ using Equation (\ref{vnsmin}).
\item {\bf Criterion~2}: The current spin-down energy of the neutron
  star in G328.4+0.2 is between $0.017\leq\dot{E_{,37}}\leq170$, where
  $\dot{E_{,37}} \equiv \dot{E}/10^{37}$~ergs.  This is based on the
  work done in \S\ref{obsres}, and is consistent with the initial
  values of $\dot{E}$ used in \S\ref{g328comp}.  Since in this
  scenario we have no estimate of the age of G328.4+0.2, we are unable
  to assume a value for $\dot{E}$ of the central neutron star and then
  calculate its initial spin-down luminosity, as we did in
  \S\ref{g328comp}. 
\item {\bf Criterion~3}: $R_{\rm pwn} < R_{\rm snr}$ for all times $t <
  t_{\rm now}$, as explained in \S \ref{g328comp}.
\item {\bf Criterion~4}: The PWN has already collided, and has been
  compressed by, the RS, as explained in \S\ref{obsres}.
\item {\bf Criterion~5}: The Central Bar created by the compression of
  the PWN is still observable.  The method of determining if this is
  satisfied is the same as the one used in \S\ref{g328comp}.
\item {\bf Criterion~6}: The PWN must have been able to form RT
  instabilities after the PWN/RS collision.  As in \S\ref{g328comp},
  we implement this requirement by requiring that $P_{\rm pwn}>P_{\rm
  snr}(R_{\rm pwn})$ for some $t>t_{\rm col}$.  Additionally, as
  explained in \S\ref{obsres}, in order for the PWN to create
  Filamentary Structure~C it must currently be unstable to R-T
  instabilities -- requiring that $P_{\rm pwn}>P_{\rm snr}(R_{\rm
  pwn})$ now.
\item {\bf Criterion~7}: As explained in \S\ref{obsres}, the
  observed Outer Protrusions in the radio require that the PWN
  currently be expanding into the SNR, $v_{\rm pwn}(t_{\rm now})>0$.
\item {\bf Criterion~8}: G328.4+0.2, must have only undergone one
  compression/re-expansion cycle.  As explained in \S\ref{obsres},
  numerical simulations of PWN inside SNRs finds that the PWN is
  disrupted after the first such cycle \citep{blondin01}.
\end{itemize}
It is important to note that, if G328.4+0.2 is a PWN, then the radio and
X-ray observations provide little information on the evolutionary phase of
the (unseen) SNR and no information on the current ratio of the PWN
and SNR radii.

Out of the 432 possible combinations of the input parameters, only
five satisfy all ten of the above criteria, as listed in
Table~\ref{pwnmodres}.  While the neutron star appears to be inside
the PWN, it is possible that this is just a projection effect.  To
evaluate the possibility that the PWN in G328.4+0.2 has already
entered the Relic PWN phase of its evolution, we calculated $v_{\rm
  ns}^{\rm min,II}$, defined as:
\begin{eqnarray}
\label{vns2eqn}
v_{\rm ns}^{\rm min,II} & = & \frac{R_{\rm G328}}{t_{\rm now}}.
\end{eqnarray}
If $v_{\rm ns}>v_{\rm ns}^{\rm min,II}$, then the G328.4+0.2 is a
Relic PWN, if not, then it is still in the Collision with the RS phase
of its evolution.

As shown in this Table, the predicted properties of G328.4+0.2 vary
substantially if G328.4+0.2 is inside a Sedov or Radiative SNR.  In
the Sedov case, G328.4+0.2 is quite young, and progenitor SN explosion
had a normal explosion energy but a low ejecta mass ($M_{\rm
  ej}\sim1M_{\odot}$), and it occurred in a low density environment.
Additionally, the neutron star formed in this explosion was spinning
rapidly, has a low surface magnetic field strength ($B_{\rm
  ns}<10^{12}$~G), and a high space velocity ($v_{\rm ns} \ga
800$~km~s$^{-1}$).  In the Radiative Case, G328.4+0.2 is substantially
older, and the progenitor SN explosion was a low kinetic energy
($E_{\rm sn} \sim 3\times10^{50}$~ergs) and high ejecta mass
expanding. The neutron star in this case was born spinning somewhat
slower and has a normal magnetic field strength for a young neutron
star. 

With the information presented in Table~\ref{pwnmodevol}, it is
possible to further refine the expected SNR and PWN properties.  As
argued in \S\ref{g328comp}, the prominence of the central bar argues
that, during the compression stage, the volume of the PWN decreased
significantly.  Though it is not possible at this time to quantify the
compression needed, an examination of Table~\ref{pwnmodevol} shows
that for only two models, ST~2 and Rad~1, did the volume of the PWN
decrease by more than 10\% -- and therefore these two models are the
most probable descriptions of G328.4+0.2.  In the case of Rad~1,
$v_{\rm ns}^{\rm min,II} \sim 100$~km~s$^{-1}$ is significantly less
then the average neutron star velocity, $v \sim 400$~km~s$^{-1}$
\citep{faucher06,hobbs05}, implying that the PWN is in the Relic PWN
phase of its evolution.  Since the sound speed inside a Radiative SNR
is quite low, $\sim100$~km~s$^{-1}$, we expect that the PWN in the
Rad~1 scenario would have a bow-shock morphology.  Since there is no
clear evidence for this in the X-ray or radio emission from
G328.4+0.2, this suggests that ST~2 is a better fit to the data. 

\begin{figure}
\begin{center}
\includegraphics[width=5.5in,angle=-90]{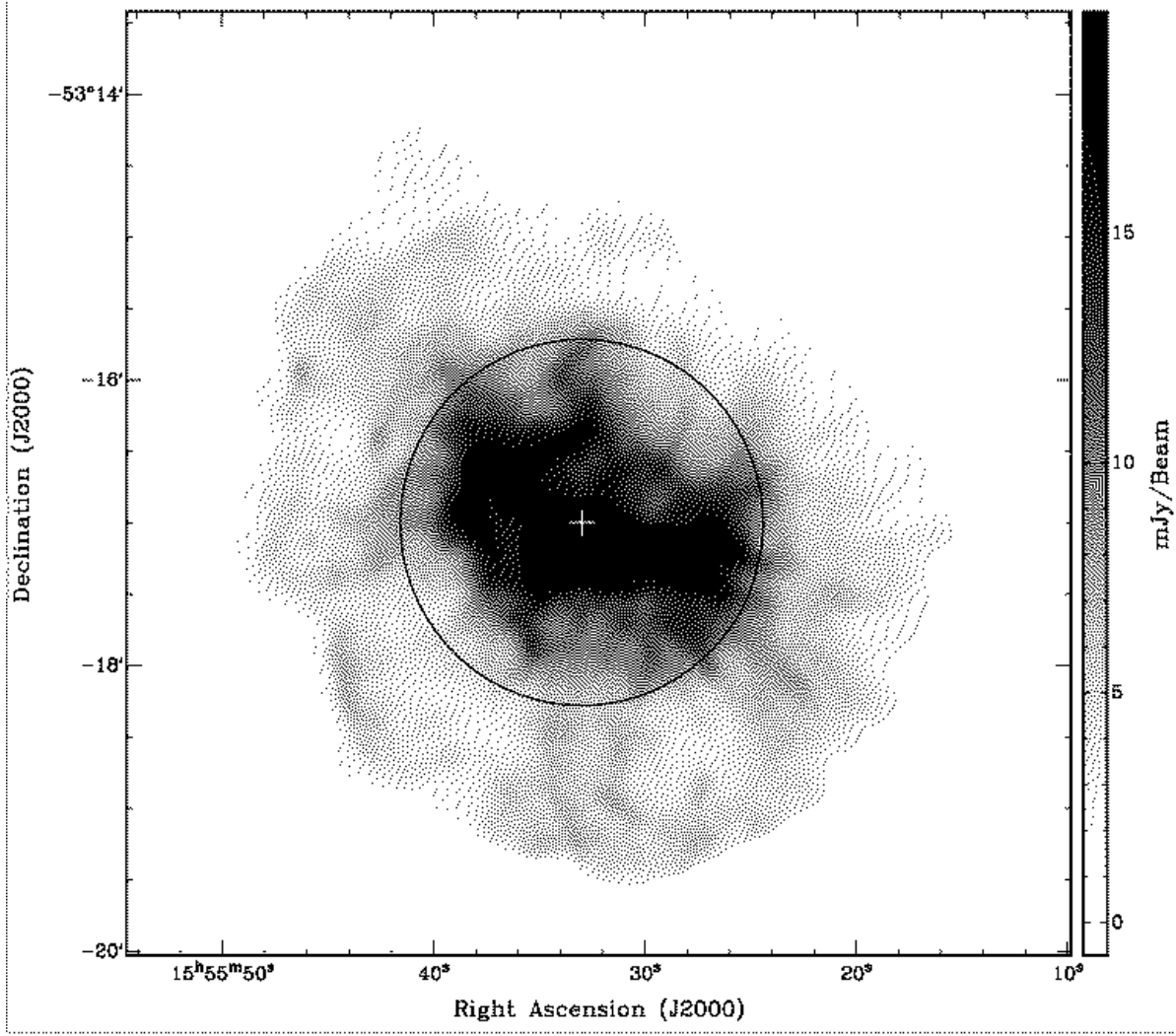}
\end{center}
\figcaption{20cm radio image of G328.4+0.2 (same data as shown in
  Fig.~\ref{radpic}), with a color scale chosen to enhance the
  visibility of Filamentary Structure~B discussed in \S\ref{radio}.
  The yellow circle indicates the size of PWN predicted in the ST~2
  model listed in Table~\ref{pwnmodevol} when it re-expanded after the
  initial compression by the SNR reverse shock, and the white cross
  indicates the center of G328.4+0.2 ($15^{\rm h}55^{\rm m}33^{\rm s},
  -53^{\circ}17^{\prime}00^{\prime \prime}$; J2000) as determined by
  \citet{gaensler00}. 
  \label{radfil}}
\end{figure}

The conclusion that ST~2 is an accurate description of G328.4+0.2 is
supported by circumstantial evidence as well.  For this scenario,
the expected radius of the termination shock around the neutron star,
$r_{\rm ts}$, defined as \citep{slane04}:
\begin{equation}
\label{rtseqn}
r_{\rm ts}^2  =  \frac{\dot{E}}{4\pi c P_{\rm pwn}},
\end{equation}
assuming a spherical wind, is $r_{\rm ts} \sim 0.6$~pc, which
corresponds to an angle of $\theta_{\rm ts} \sim 8d_{17}\arcsec$ --  a
distance which is comparable to the offset between Clump~1 and Clump~2
derived in \S\ref{image}.  Another interesting feature for this model
is that, as shown in Fig.~\ref{radfil}, the radius of the PWN at
the time of re-expansion is similar to that of the outer parts of the
central filamentary structures discussed in \S\ref{radio}.  While this
correlation might be coincidental, this could imply that the
hydrodynamic instabilities formed at the PWN/SNR interface during
the re-expansion disrupted the shell of material swept up by the PWN
-- consistent with the simulation of \citet{blondin01}.  While not
definitive, these two pieces of evidence argue that ST~2 is a
reasonable description of G328.4+0.2.

\begin{figure}
\begin{center}
\includegraphics[width=4.0in,angle=90]{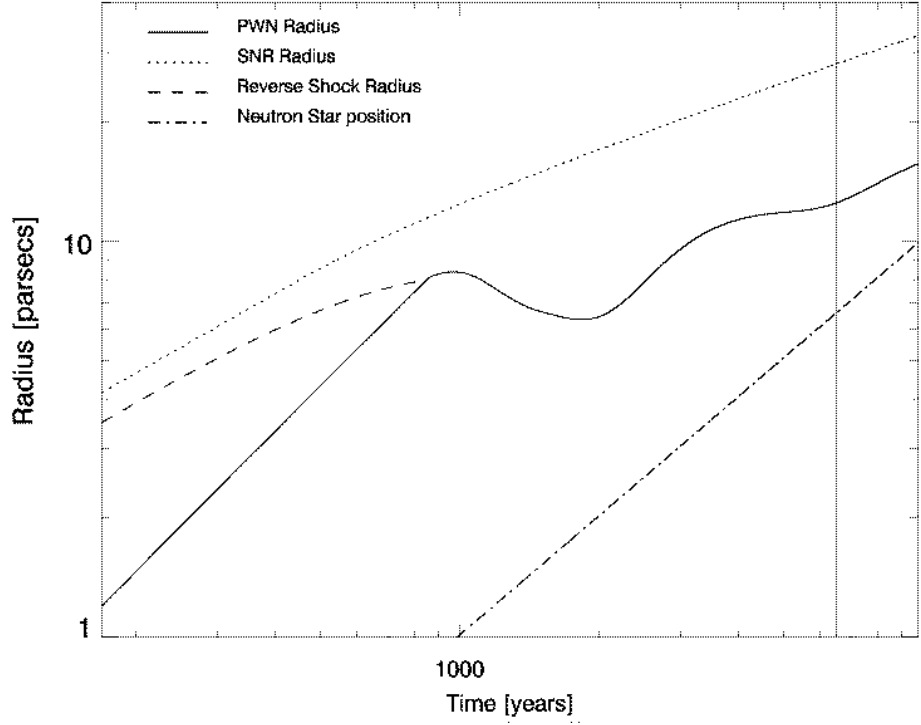}
\end{center}
\figcaption{The radius of the PWN, SNR, and SNR reverse shock as well
  as the location of the neutron star as a function of time for the
  favored (ST~2; Table \ref{pwnmodres}) scenario if G328.4+0.2 is a PWN.
  The vertical line indicates the current age of the system. \label{st2evol}}
\end{figure}

The properties of ST~2 are given in Table \ref{st2prop}, and the
evolution of $R_{\rm pwn}$ is shown in Fig.~\ref{st2evol}.  It is
interesting to note that in this scenario, the PWN collides with the
RS at a time $t_{\rm col}<\tau_0$ ($t_{\rm col} \approx 850$~years) so
energy injection by the neutron star into the PWN after the PWN/RS
collision is important to the PWN's evolution during this stage.  It
is important to note that this model predicts that the neutron star
powering G328.4+0.2 is the most energetic neutron star in the Milky
Way, as well as one of the fastest.  Given that G328.4+0.2 is largest
and has the highest radio luminosity of any known PWN, it is not
surprising that it was formed by such a powerful neutron star.
Finally, the age and $\dot{E}$ predicted by this method are similar to
those predicted by \citet{gaensler00}.

\begin{figure}
\begin{center}
\includegraphics[width=3.4in,angle=0]{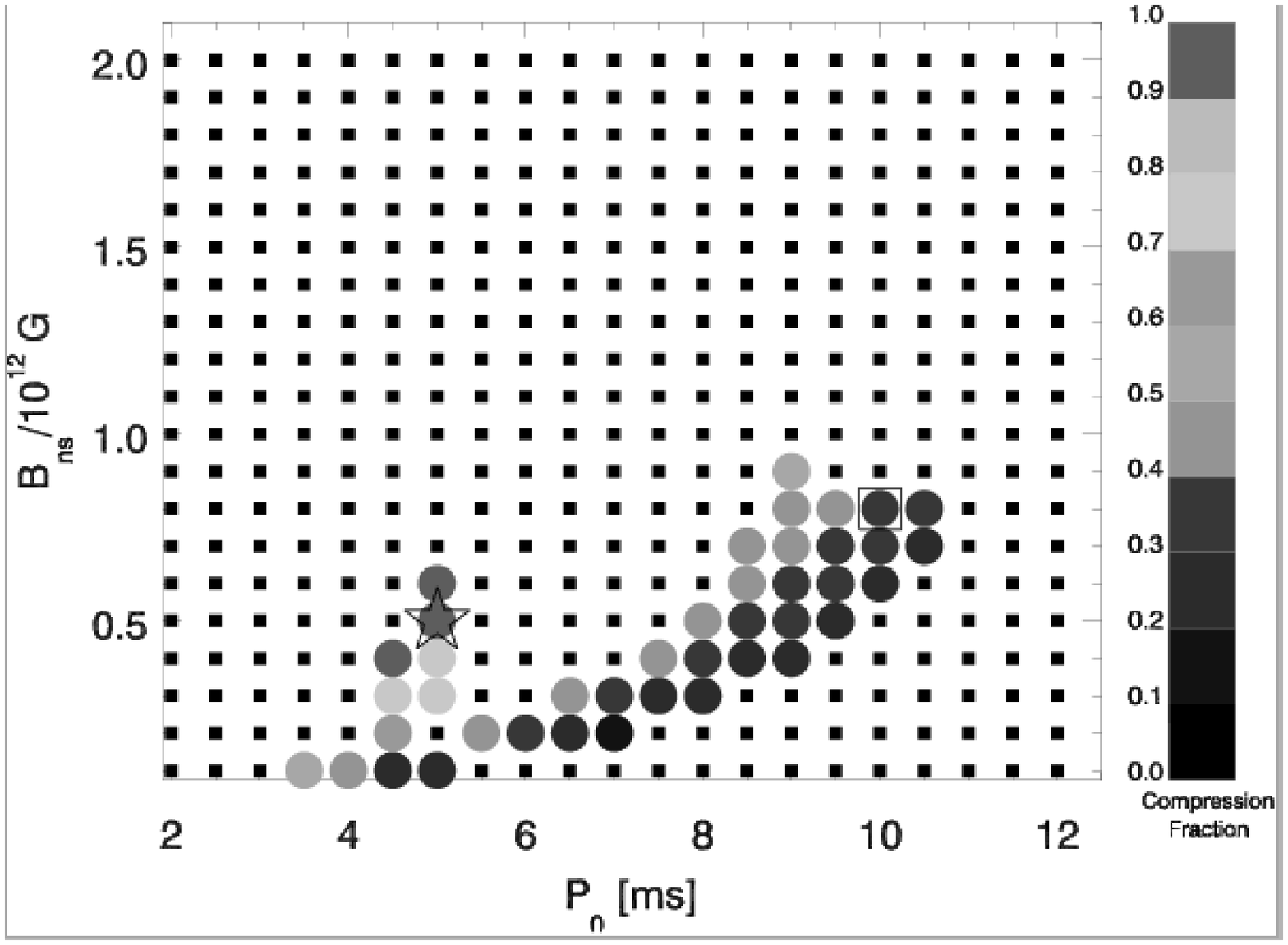}
\includegraphics[width=3.4in,angle=0]{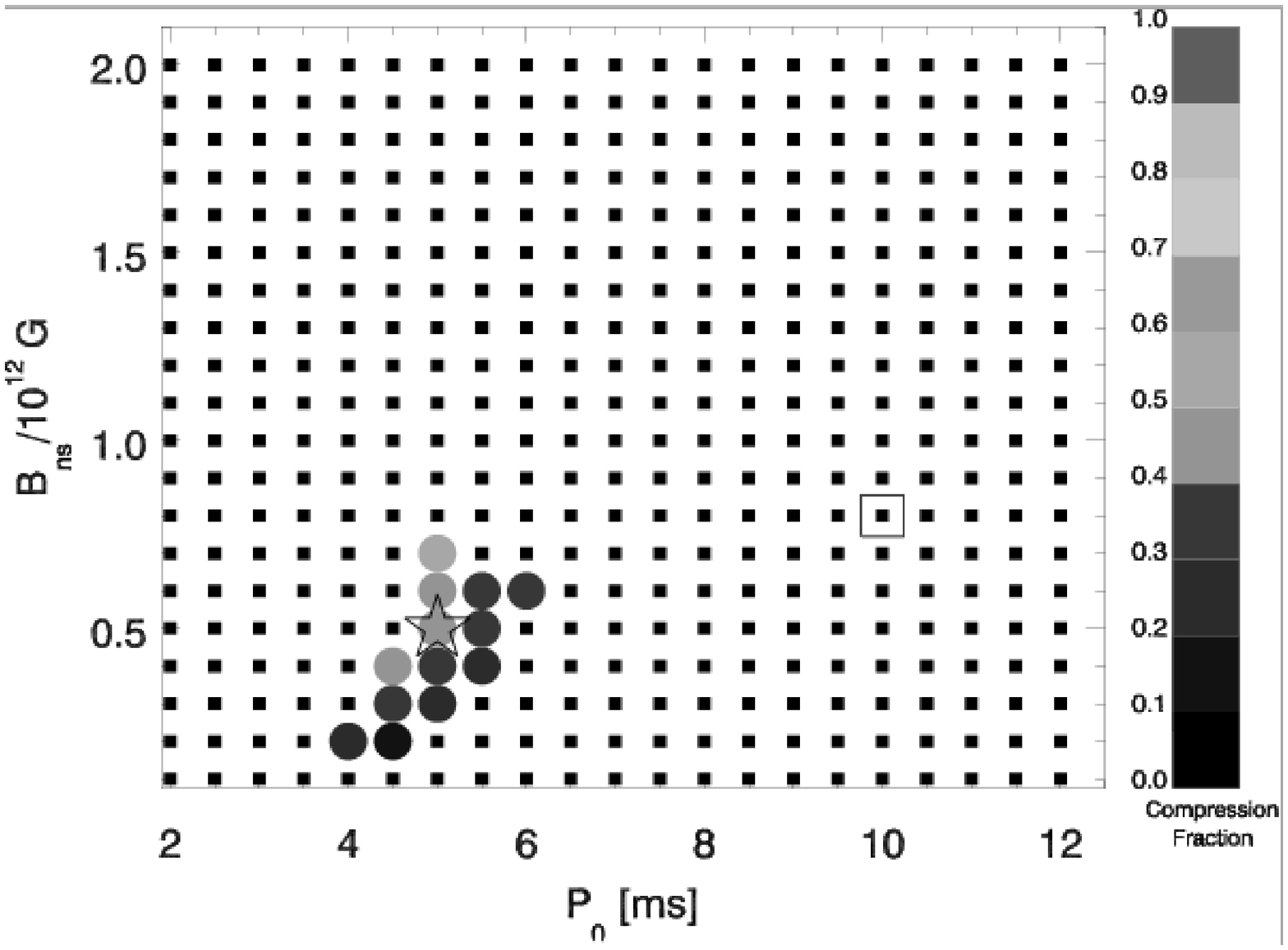}
\includegraphics[width=3.4in,angle=0]{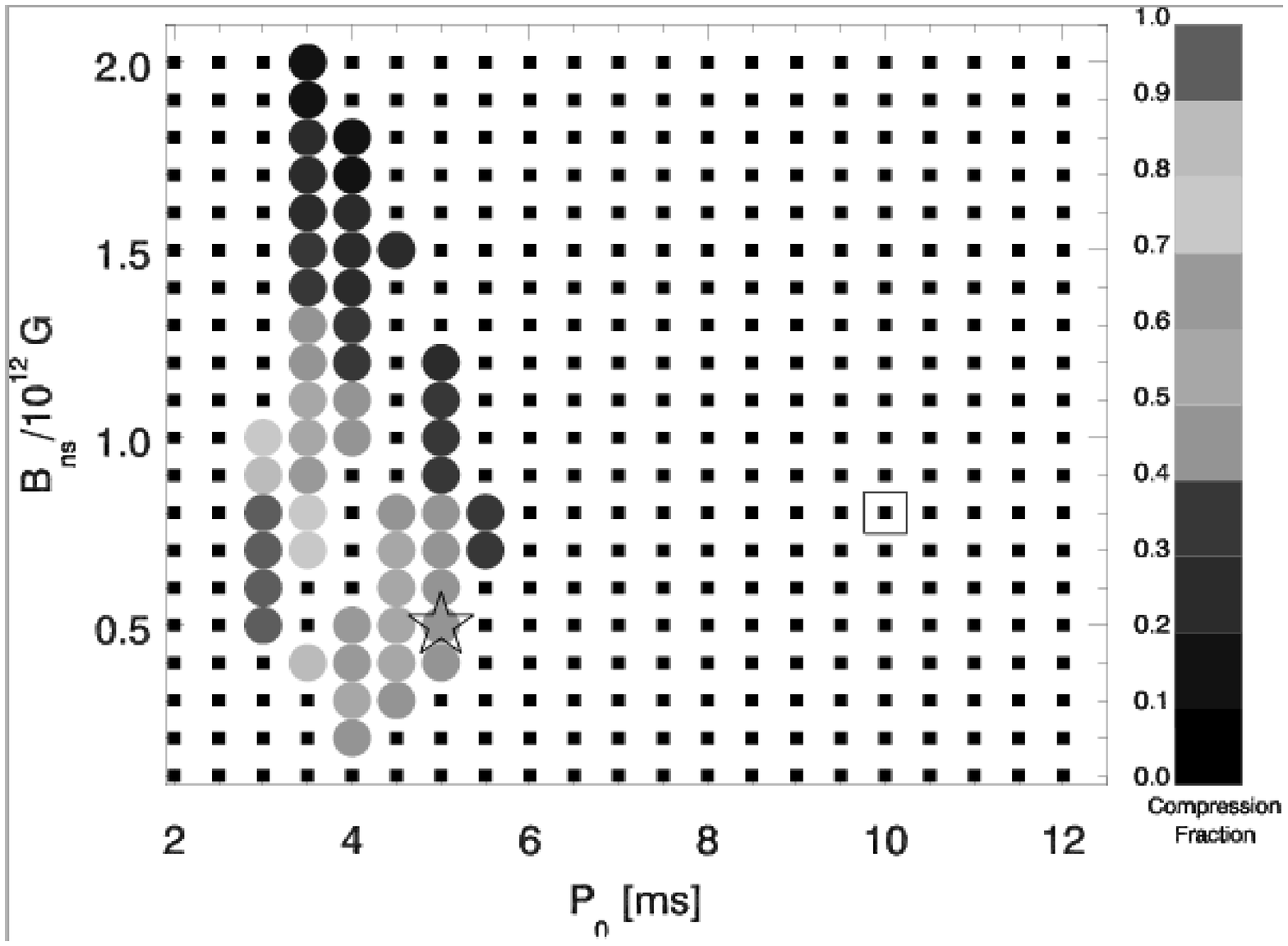}
\end{center}
\figcaption{The results for varying $P_0$ and $B_{\rm ns}$ for a 
  $E_{\rm sn}=10^{51}$~ergs, $M_{\rm ej}=1 M_{\odot}$ ({\it top}),
  $E_{\rm sn}=3\times10^{51}$~ergs, $M_{\rm ej}=1 M_{\odot}$ ({\it
  middle}), and $E_{\rm sn}=4\times10^{51}$~ergs, $M_{\rm ej}=3.25 
  M_{\odot}$ ({\it bottom}) SN explosion, assuming $n=0.03$~cm$^{-3}$.
  The small black squares indicate models which failed the criteria
  described in \S\ref{g328pwn}, while the colored circles indicate
  scenarios which passed.  The color represents the Compression
  Fraction of the PWN, defined as the ratio of the PWN's volume at the
  beginning and end of the compression stage.  The Compression
  Fraction of the ST~2 case given in Table \ref{pwnmodres} is 0.44
  (light blue on this color scale), and lower values correspond to a
  more substantial compression.  The star indicates the position of a
  $P_0=5$~ms, $B_{\rm ns}=5\times10^{11}$~G neutron star, while the
  large square indicates the position of a $P_0=10$~ms, $B_{\rm
  ns}=8\times10^{11}$~G neutron star -- the two neutron stars used in
  Fig. \ref{varysn}.
  \label{varyns}}  
\end{figure}

In order to better understand the limitations of the approach in
determining the properties of the neutron star and SNR in G328.4+0.2, 
we have run the model presented in \S\ref{model} over a finer grid of
parameters and evaluated the resulting PWN evolution using the same
criteria as above.  In Fig.~\ref{varyns}, we show which values of
$P_0$ and $B_{\rm ns}$ pass all of this criteria for three different
kind of SN explosions: $E_{\rm sn}=10^{51}$~ergs and 
$M_{\rm ej}=1~M_\odot$, $E_{\rm sn}=3\times10^{51}$~ergs and 
$M_{\rm ej}=1~M_\odot$, and $E_{\rm sn}=4\times10^{51}$~ergs and
$M_{\rm ej}=3.25~M_{\odot}$, assuming an ambient density with
$n=0.03$.  The first set of SN parameters corresponds to ST~1, the second
to ST~2, and third to a higher ejecta mass SN explosion is compatible
with a neutron star with the same parameters as ST~2.  For the first
case, we find that a wide range of $P_0$ values are allowed but that
$B_{\rm ns} \la 10^{12}$~G.  In fact, for this set of SN parameters a
$P_0 \sim 10$~ms, $B_{\rm ns} \sim 8\times10^{11}$~G neutron star
results in a PWN which is compressed a similar amount as in ST~2.  In
the second case, we find that $P_0$ and $B_{\rm ns}$ are tightly
constrained around $P_0 \approx 5$~ms and $B_{\rm ns} \approx
5\times10^{11}$~G.  In the third set of SN parameters, we find that
$P_0 \la 6$~ms, but that $B_{\rm ns}$ spans a wide range of values,
$\sim10^{11}-2\times10^{12}$~G.

To determine the allowed values of $E_{\rm sn}$ and $M_{\rm ej}$, we
followed the same procedure as above using two different sets of
neutron star parameters: $P_0=5$~ms and $B_{\rm ns}=5\times10^{11}$~G 
(the neutron star parameters in the ST~1 and ST~2 scenarios), and
$P_0=10$~ms and $B_{\rm ns}=8\times10^{11}$~G.  For the first case, 
only models with $E_{\rm sn} \sim 1-4\times10^{51}$~ergs and $M_{\rm
  ej} \sim 0.5-3.5 M_{\odot}$ satisfy the criteria above --  though a
substantial compression of the PWN requires $E_{\rm sn} \ga
2\times10^{51}$~ergs.  In the second case, we find that only models
with $E_{\rm sn} \la 10^{51}$~ergs and $M_{\rm ej} \sim 0.5-3.5
M_{\odot}$ are allowed.

While this error analysis shows that the method used above to
determine the properties of the neutron star and SN explosion which
formed G328.4+0.2 is unable to do so to much better than an order of
magnitude, the different combinations values of $P_0$, $B_{\rm ns}$,
$E_{\rm sn}$, and $M_{\rm ej}$ which are allowed predict different
physical properties for G328.4+0.2 which are testable with further
observations.  For example, in the case of $P_0=10$~ms, 
$B_{\rm ns}=8\times10^{11}$~G, $E_{\rm sn}=1\times10^{51}$~ergs, and 
$M_{\rm ej}=1~M_{\odot}$, G328.4+0.2 is $\sim13,000$~years old, twice
the age predicted in the ST~2 model as shown in Table~\ref{st2prop},
and as a result the required velocity of the neutron star is
significantly lower, $v_{\rm ns}^{\rm min} \sim 500$~km~s$^{-1}$.  The
predicted period for the neutron star in this scenario is also
significantly slower than required by the ST~2 scenario, $P\sim24$~ms,
with a value of $\dot{E}$ approximately an order of magnitude lower
than that in the ST~2 scenario.  The termination shock radius for this
set of parameters is $\sim5\arcsec$, considerable smaller than the
$\sim8\arcsec$ for the ST~2 scenario and detectable with the {\it
  Chandra} X-ray Observatory.

\section{Conclusions}
\label{conclusion}
In this paper, we first presented new X-ray (\S\ref{xray}) and radio
(\S~\S\ref{radio}, \ref{pulsar}) and observations of Galactic
non-thermal radio and X-ray source, G328.4+0.2, from which we infer
the current properties and evolutionary history of this source
(\S\ref{obsres}).  We then presented a simple hydrodynamic model for
the evolution of a PWN inside a SNR (\S\ref{model}), which is used to
determine which values of $E_{\rm sn}$, $M_{\rm ej}$, $n$, $P_0$, and
$B_{\rm ns}$ are able to reproduce the properties discussed in
\S\ref{obsres} if G328.4+0.2 was a Composite SNR (\S\ref{g328comp}) or
a PWN (\S\ref{g328pwn}).  As a result of this analysis, we determine
the G328.4+0.2 is a PWN inside an undetected SNR.  Though we are not
able to precisely determine the properties of the SN explosion and the
neutron star which have created this system, our analysis implies that
the neutron star in G328.4+0.2 was born with an initial period $P_0
\la 10$~ms, has a lower than average surface dipole magnetic field
strength, and has a higher than average spatial velocity $v_{\rm ns}
\ga 400$~km~s$^{-1}$.  We assume determining that the SN explosion
which created the neutron star had a normal explosion energy, $E_{\rm
sn} \sim 10^{51}$~ergs, but a relative low ejecta mass, $M_{\rm ej}
\la 4M_{\odot}$.  Future X-ray and radio observations can
significantly decrease this uncertainty, particularly if they are able
to either detect pulsations from the neutron star or continuum X-ray
or radio emission from the currently undetected SNR in this system.

While we are not able to definitely determine the initial period
($P_0$) or surface magnetic field strength ($B_{\rm ns}$) of the
neutron star, nor the kinetic energy ($E_{\rm sn}$) or ejecta mass
($M_{\rm ej}$) of the progenitor SN explosion, the estimates quoted
above are of interest.  Our non-detection of the pulsar via radio
pulsations is not particularly constraining, due to the very large
distance of the PWN.  The low magnetic field but rapid initial period
predicted for the neutron star in G328.4+0.2 has implications for
models concerning the origin of neutron star magnetic fields.  For
example, according to the $\alpha-\Omega$ dynamo model of
\citep{thompson93}, neutron star born spinning rapidly ($P_0 \la
5$~ms) should have a strong dipole component to their surface magnetic
fields ($B_{\rm ns} \gg 10^{12}$~G).  If the ST~2 scenario proves to
be correct, then the low magnetic field of the neutron star in this
system would be a problem for such a model.  Additionally, the low
ejecta mass inferred in this scenario requires that the progenitor of
this system was either a single, massive star ($M \ga 35 M_{\odot}$)
which exploded in a Type Ib/c SN \citep{woosley95}, or was initially
in a binary system.

Finally, the method used in this paper to study G328.4+0.2 is
complementary to other methods used (e.g. \citeauthor{chevalier05}
\citeyear{chevalier05}) to infer the initial period and magnetic field
strength of other neutron stars in young PWN as well as the properties
of the SN explosion in which they were formed, and is easily
applicable to other such systems.

\acknowledgements 

JDG would like to thank Niccolo Bucciantini, Shami Chatterjee, Roger
Chevalier, Tracey DeLaney, David Kaplan, Kelly Korreck, Cara Rakowski,
and John Raymond for useful discussions, and the anonymous referee for
many useful comments.  We are extremely grateful to John Reynolds for
prompt scheduling and observing assistance at Parkes in 2007.  The
Australia Telescope is funded by the Commonwealth of Australia for
operation as a National Facility managed by CSIRO. JDG and BMG were
supported in this work by {\it XMM} grant NAG5-13202 and LTSA grant
NAG5-13032.

\bibliography{ms}
\bibliographystyle{apj}

%% Tables
\newpage
\begin{table}
\begin{center}
\caption{Spatial components of the X-ray emission from G328.4+0.2
  \label{sherpares}}
\begin{tabular}{ccc}
\hline
\hline
{\sc Component} & {\sc Parameter} & {\sc Value} \\
\hline
{\bf background} & constant & $0.09^{+0.06}_{-0.07}$ \\
\hline
{\bf Clump 1} & $r_0$ & $1\farcs1_{-0\farcs7}^{+1\farcs5}$ \\
         & Position & 
             $15^{\rm h}55^{\rm m}26\fs68^{+0\fs04}_{-0\fs05}$,
             $-53\degr18\arcmin02\farcs7^{+0\farcs6}_{-0\farcs6}$ \\ 
         & $A$ & $6.6^{+27.6}_{-3.2}$ \\
         & $\alpha$ & $0.9^{+1.0}_{-0.3}$ \\
\hline
{\bf Clump 2} & FWHM & $19\arcsec^{+10\arcsec}_{-10\arcsec}$ \\
         & Position & 
             $15^{\rm h}55^{\rm m}27\fs5^{+0\fs2}_{-0\fs1}$,
             $-53\degr17\arcmin54\farcs6^{+3\farcs4}_{-3\farcs8}$ \\ 
         & $e$ & $0.4^{+0.2}_{-0.3}$ \\
         & $\theta$ & $280\degr^{+20\degr}_{-80\degr}$ \\
         & $A$ & $0.8^{+0.3}_{-0.2}$ \\
\hline
{\bf Diffuse} & FWHM & $67^{+18}_{-11}$ \\
         & Position & 
             $15^{\rm h}55^{\rm m}26\fs6^{+0\fs2}_{-0\fs2}$,
             $-53\degr17\arcmin48\farcs3^{+5\arcsec}_{-4\arcsec}$ \\ 
         & $e$ & $0.4^{+0.1}_{-0.1}$ \\
         & $\theta$ & $130\degr^{+10\degr}_{-10\degr}$ \\
         & $A$ & $0.4^{+0.1}_{-0.1}$ \\
\hline
\hline
\end{tabular}
\end{center}
{\sc Note.} -- Results from the spatial fit to the X-ray emission from
  G328.4+0.2, as described in \S \ref{image}.  Fitting was done using
  the {\sc Sherpa} software package, and the errors reflect the 90\%
  confidence level.  Clump~1 was fitted to a 2D Lorentzian, defined as
  $f(r)=A \left(1+\frac{r^2}{r_0^2}\right)^{-\alpha}$, where $f(r)$ is
  the expected number of counts at radius $r$ away from the center of
  the source, $A$ is given in counts, and the core-radius $r_0$ is in
  arc-seconds. The background component is given in counts
  pixel$^{-1}$.  Both Clump~2 and Diffuse were modeled with elliptical
  2D Gaussians, where the full width, half maximum (FWHM) is given in
  arc-seconds, the ellipticity is $e$, defined as $1-b/a$, where $b$
  and $a$ are, respectively, the major and minor axis of the source,
  the position angle $\theta$ is given in degrees counterclockwise from
  north, and the amplitude $A$ is given in counts.
\end{table}

\begin{table}
\begin{center}
\caption{Spectral Fits to {\sc Mos1 + Mos2 + pn} Data \label{specres}}
\begin{tabular}{c|cccc}
\hline
\hline
{\sc Parameter} & \multicolumn{4}{c}{Value} \\
\hline
\hline
Model & {\bf phabs * pow} & {\bf phabs * bbodyrad} & {\bf phabs *
  bremss} & {\bf phabs * ray} \\
\hline
N$_{H}$ & $11.6_{-2}^{+3}\times10^{22}$ &
$6.4_{-1}^{+1}\times10^{22}$ & $10.4_{-1}^{+2}\times10^{22}$ &
$7.8^{+1}_{-1}\times10^{22}$ \\
$\Gamma$ or kT & $2.0^{+0.4}_{-0.2}$ & $1.6^{+0.1}_{-0.1}$ &
$9.5^{+5}_{-3}$ & $62.9^{+ >900}_{-30}$ \\
Absorbed Flux & $4.6_{-3}^{+1}\times10^{-13}$ &
$4.4_{-0.5}^{+0.1}\times10^{-13}$ & $4.5_{-1.5}^{+0.4}\times10^{-13}$
& $4.9_{-2.6}^{+0.3}\times10^{-13}$ \\ 
Unabsorbed Flux & $1.9_{-0.9}^{+0.1}\times10^{-12}$ &
$6.5_{-1.0}^{+0.1}\times10^{-13}$ & $1.2_{-0.5}^{+0.1}\times10^{-12}$
& $9.6_{-4.2}^{+0.1}\times10^{-13}$ \\ 
\hline
$\chi^2$ & 103.3/131 & 98.4/131 & 101.7/131 & 122.8/131 \\
Reduced $\chi^2$ & 0.81/128 & 0.77/128 & 0.79/128 & 0.96/128 \\
\hline
\hline
\end{tabular}
\end{center}
{\sc Note} -- Results from joint fits to the {\sc mos1}, {\sc mos2}, and
  {\sc pn} spectrum between 0.5 and 10~keV. The model {\bf phabs *
  pow} refers to a power-law attenuated by interstellar absorption,
  {\bf phabs * bbodyrad} refers to a blackbody attenuated by
  interstellar absorption, {\bf phabs * bremss} refers to a
  Bremsstrahlung source spectrum attenuated by interstellar
  absorption, and {\bf phabs * ray} refers to a Raymond-Smith thermal
  plasma spectrum attenuated by interstellar absorption.  In the
  table, N$_H$ is given in cm$^{-2}$, kT in keV, and flux in
  ergs~cm$^{-2}$~s$^{-1}$, both the absorbed and unabsorbed flux are
  calculated between 0.5 and 10~keV, and the errors represent the 90\%
  confidence level.
\end{table}

\begin{table}
\begin{center}
\caption{Expected Properties of G328.4+0.2 if it is a Composite
  SNR \label{modg328comp}}
\begin{tabular}{cc|cc|cc}
\hline
\hline
\multicolumn{2}{c}{\it Supernova Remnant Properties} & 
\multicolumn{2}{c}{\it Pulsar Wind Nebula Properties} & 
\multicolumn{2}{c}{\it Neutron Star Properties} \\
\hline
\hline
{\sc Parameter} & {\sc Value} & {\sc Parameter} & {\sc Value} & {\sc
  Parameter} & {\sc Value} \\
\hline
${\bf E_{\bf sn}}$ & ${\bf 1\times10^{51}}$~{\bf ergs} & $E_{\rm pwn}$ &
$4.8\times10^{50}$~ergs & $\dot{E_0}$ &
$2.5\times10^{40}$~ergs~s$^{-1}$ \\
${\bf M_{\bf ej}}$ & {\bf 1}~${\bf M_{\odot}}$ & 
$R_{\rm pwn}$ & 8.7~parsecs & $P_0$ & 3.8~ms \\
Phase & Sedov-Taylor & $M_{\rm sw}^{\rm pwn}$ & 5.3~$M_{\odot}$ &
${\bf \tau_0}$ & {\bf 1730~years} \\
$P_{\rm snr}(R_{\rm pwn})$ & $1.4\times10^{-9}$~dynes & $P_{\rm pwn}$
& $2.0\times10^{-9}$~erg~cm$^{-4}$ & $B_{\rm ns}$ & $1.1\times10^{12}$~G \\
$v_{\rm ej}(R_{\rm pwn})$ & 400~km~s$^{-1}$ & $v_{\rm pwn}$ &
700~km~s$^{-1}$ & $v_{\rm ns}^{\rm min}$ & 1300~km~s$^{-1}$ \\
$t_{\rm now}$ & 4900~years & $r_{\rm ts}$ & 0.5~parsecs &
${\bf \dot{E}}$ & $\bf{1.7\times10^{39}}$~{\bf ergs~s}$^{\bf -1}$ \\
${\bf n}$ & $\bf{0.32}$~{\bf cm}$^{\bf -3}$ & $\theta_{\rm ts}$ &
$6\farcs7$ & $P$ & 7.5~ms \\ 
$\cdots$ & $\cdots$ & $\cdots$ & $\cdots$ & $\dot{P}$ &
$1.8\times10^{-14}$ s/s \\ 
$\cdots$ & $\cdots$ & $\cdots$ & $\cdots$ & $\tau_c$ &
6700~years \\
$\cdots$ & $\cdots$ & $\cdots$ & $\cdots$ & $E_{\rm psr}$ &
$1.0\times10^{51}$~ergs \\
\hline
\hline
\end{tabular}
\end{center}
{\sc Note.} -- The values in {\bf bold} are model assumptions, while the
  others are predicted by the model presented in
  \S\ref{model}. $P_{\rm snr}(R_{\rm pwn})$ is the pressure inside of 
  the SNR just outside of the PWN, $v_{\rm ej}(R_{\rm pwn})$ is the
  velocity of material inside the SNR just outside of the PWN, $E_{\rm
  pwn}$ is the internal energy of the PWN, $R_{\rm pwn}$ is the radius
  of the PWN, $M_{\rm sw}^{\rm pwn}$ is the mass of material swept up
  by the PWN, $P_{\rm pwn}$ is the internal pressure of the PWN,
  $v_{\rm pwn}$ is the expansion velocity of the PWN, $r_{\rm ts}$ is
  the radius of the termination shock around the neutron star inside
  the PWN, calculated using Equation \ref{rtseqn}, $\theta_{\rm ts}$
  is the angular size of this feature assuming $d_{17} \equiv 1$,
  $B_{\rm ns}$ is the dipole magnetic field of the neutron star in
  G328.4+0.2 according to this model, $P$ is the period of the neutron
  star in G328.4+0.2, $\dot{P}$ is the period-derivative of the
  neutron star,  $\tau_c$ is the characteristic age of the neutron
  star, defined as $\tau_c=P/(2\dot{P})$, and $E_{\rm psr}$ is the
  total amount of energy injected by the neutron star into G328.4+0.2
  for $t<t_{\rm now}$.  All values are given for $t=t_{\rm now}$
  unless otherwise noted.
\end{table}

\begin{table}
\begin{center}
\caption{Scenarios for G328 as a PWN inside an 
  Undetected SNR \label{pwnmodres}}
\begin{tabular}{cccccccccccc}
\hline
\hline
Scenario \# & $\tau_0$ & $P_0$ & $E_{\rm sn,51}$ & $M_{\rm ej}$ & $n$
& $t_{\rm now}$ & SNR~Phase & $R_{\rm snr}$ & $B_{\rm ns,12}$ &
$v_{\rm ns}^{\rm min}$ & $v_{\rm ns}^{\rm min,II}$ \\
\hline
ST~1 & 1730 & 5.0 & 1.00 & 1 & 0.03 & 5100 & Sedov-Taylor & 19.8 & 0.5
& 1300 & 2400 \\ 
ST~2 & 1730 & 5.0 & 3.16 & 1 & 0.03 & 6500 & Sedov-Taylor & 28.0 & 0.5
& 1000 & 1900 \\
\hline
Rad~1 & 430 & 10.0 & 0.32 & 10 & 0.32 & 84200 & Radiative & 28.4 & 2.0 & 100 & 
100 \\
Rad~2 & 770 & 10.0 & 0.32 & 10 & 0.32 & 52400 & Radiative & 24.8 & 1.5 & 100 & 
200 \\
Rad~3 & 770 & 10.0 & 0.32 & 10 & 1.00 & 101400 & Radiative & 22.1 &
1.5 & 100 & 100 \\
\hline
\hline
\end{tabular}
\end{center}
{\sc Note.} -- Values for $\tau_0$, $P_0$, $E_{\rm sn}$, $M_{\rm ej}$,
 and $n$ that satisfy the criteria listed in \S\ref{g328pwn} for G328
 being a PWN inside an undetected SNR.  The value of $\tau_0$ is given
 in years, $P_0$ in ms, $E_{\rm sn,51}\equiv E_{\rm sn}/10^{51}$~ergs,
 $M_{\rm ej}$ in solar masses, $n$ in cm$^{-3}$, $t_{\rm now}$ in
 years, $R_{\rm snr}$ in parsecs, $B_{\rm ns}=B_{\rm
 ns,12}\times10^{12}$ ~G, $v_{\rm ns}^{\rm min}$ is in km~s$^{-1}$,
 and $v_{\rm ns}^{\rm min,II}$ is also in km~s$^{-1}$.
\end{table}

\begin{table}
\begin{center}
\caption{Compression/Expansion properties of G328.4+0.2 
 if it is a PWN \label{pwnmodevol}}
\begin{tabular}{c|cccc}
\hline
\hline
Scenario \# & $t(v_{\rm pwn}=0)$ & Central Bar Lifetime & $R_{\rm
  pwn}(v_{\rm pwn}=0)$ & $\frac{V_{\rm pwn}^{\rm re-exp}}{V_{\rm
    pwn}^{\rm compres}}$ \\
\hline
ST~1 & 1648.5,~2432.0 & 63930 & 8.20,~7.96 & 0.91 \\
ST~2 & 969.0,~1837.0 & 35250 & 8.36,~6.34 & 0.44 \\
\hline
Rad~1 & 12772.0,~26541.0 & 243480 & 8.53,~7.55 & 0.69 \\
Rad~2 & 13336.5,~21112.5 & 256230 & 8.48,~8.27 & 0.93 \\
Rad~3 & 9761.5,~11035.0 & 102640 & 5.72,~5.72 & 1.00 \\
\hline
\hline
\end{tabular}
\end{center}
{\sc Note.} -- In this table, values Scenario \# correspond to 
 the values of $\tau_0$, $P_0$, $E_{\rm sn}$, $M_{\rm ej}$, and $n$ 
 given in Table \ref{pwnmodres}.  $t(v_{\rm pwn}=0)$ and the 
 Central Bar lifetime are given in years, and 
 $R_{\rm pwn}(v_{\rm pwn}=0)$ is given in parsecs.
 $\frac{V_{\rm pwn}^{\rm re-exp}}{V_{\rm pwn}^{\rm compres}}$ is 
 the ratio of the volume of the PWN at re-expansion and 
 compression.
\end{table}

\begin{table}
\begin{center}
\caption{Expected Properties of G328.4+0.2 if it is a PWN (ST~2
  scenario in Table \ref{pwnmodres})\label{st2prop}}
\begin{tabular}{cc|cc|cc}
\hline
\hline
\multicolumn{2}{c}{\it Supernova Remnant Properties} & 
\multicolumn{2}{c}{\it Pulsar Wind Nebula Properties} & 
\multicolumn{2}{c}{\it Neutron Star Properties} \\
\hline
\hline
{\sc Parameter} & {\sc Value} & {\sc Parameter} & {\sc Value} & {\sc
  Parameter} & {\sc Value} \\
\hline
${\bf E_{\bf sn}}$ & ${\bf 3.2\times10^{51}}$~{\bf ergs} & $E_{\rm pwn}$ &
$3.2\times10^{50}$~ergs & $\dot{E_0}$ &
$1.5\times10^{40}$~ergs~s$^{-1}$ \\
${\bf M_{\bf ej}}$ & ${\bf1~M_{\odot}}$ & $R_{\rm pwn}$ & 12.5~parsecs
& ${\bf P_0}$ & {\bf 5~ms} \\
Phase & Sedov-Taylor & $M_{\rm sw}^{\rm pwn}$ & 0.4~$M_{\odot}$ &
$\bf \tau_0$ & {\bf 1730~years} \\
$P_{\rm snr}(R_{\rm pwn})$ & $3.6\times10^{-10}$~dynes & $P_{\rm pwn}$
& $4.4\times10^{-10}$~dynes & $B_{\rm ns}$ & $5\times10^{11}$~G \\
$v_{\rm ej}(R_{\rm pwn})$ & 460~km~s$^{-1}$ & $v_{\rm pwn}$ &
790~km~s$^{-1}$ & $v_{\rm ns}^{\rm min}$ & 990~km~s$^{-1}$ \\
Age ($t_{\rm now}$) & 6500~years & $r_{\rm ts}$ & 0.6~parsecs &
$\dot{E}$ & $6.4\times10^{38}$~ergs~s$^{-1}$ \\
$\bf n$ & $\bf 0.03$~{\bf cm}$^{\bf -3}$ & $\theta_{\rm ts}$ &
$7\farcs7$ & $P$ & 10.9~ms \\ 
$\cdots$ & $\cdots$ & $\cdots$ & $\cdots$ & $\dot{P}$ &
$2.1\times10^{-14}$ s/s \\ 
$\cdots$ & $\cdots$ & $\cdots$ & $\cdots$ & $\tau_{\rm c}$ &
8200~years \\
$\cdots$ & $\cdots$ & $\cdots$ & $\cdots$ & $E_{\rm psr}$ &
$6.3\times10^{50}$~ergs \\
\hline
\hline
\end{tabular}
\end{center}
{\sc Note.} -- The model assumptions are given in {\bf bold}. $P_{\rm
  snr}(R_{\rm pwn})$ is the pressure inside just outside of the PWN,
  $v_{\rm ej}(R_{\rm pwn})$ is the velocity of material inside just
  outside of the PWN, $E_{\rm pwn}$ is the internal energy of the PWN,
  $R_{\rm pwn}$ is the radius of the PWN, $M_{\rm sw}^{\rm pwn}$ is
  the mass of material swept up by the PWN, $P_{\rm pwn}$ is the
  internal pressure of the PWN, $v_{\rm pwn}$ is the expansion
  velocity of the PWN, $r_{\rm ts}$ is the radius of the termination
  shock around the neutron star inside the PWN, $\theta_{\rm ts}$ is
  the predicated angular radius of this feature assuming $d_{17}
  \equiv 1$, $B_{\rm ns}$ is the predicted dipole magnetic field of
  the neutron star in G328.4+0.2, $P$ is the predicted period of the
  neutron star in G328.4+0.2, $\dot{P}$ is the predicted
  period-derivative of the neutron star, $\tau_c$ is characteristic
  age of the neutron star, and $E_{\rm psr}$ is the total amount of
  energy injected by the neutron star into G328.4+0.2 for $t<t_{\rm
  now}$.  All values are given for $t=t_{\rm now}$ unless otherwise
  noted.
\end{table}

\newpage
\clearpage
%% Figures

\begin{figure}
\begin{center}
\includegraphics[width=5.0in,angle=0]{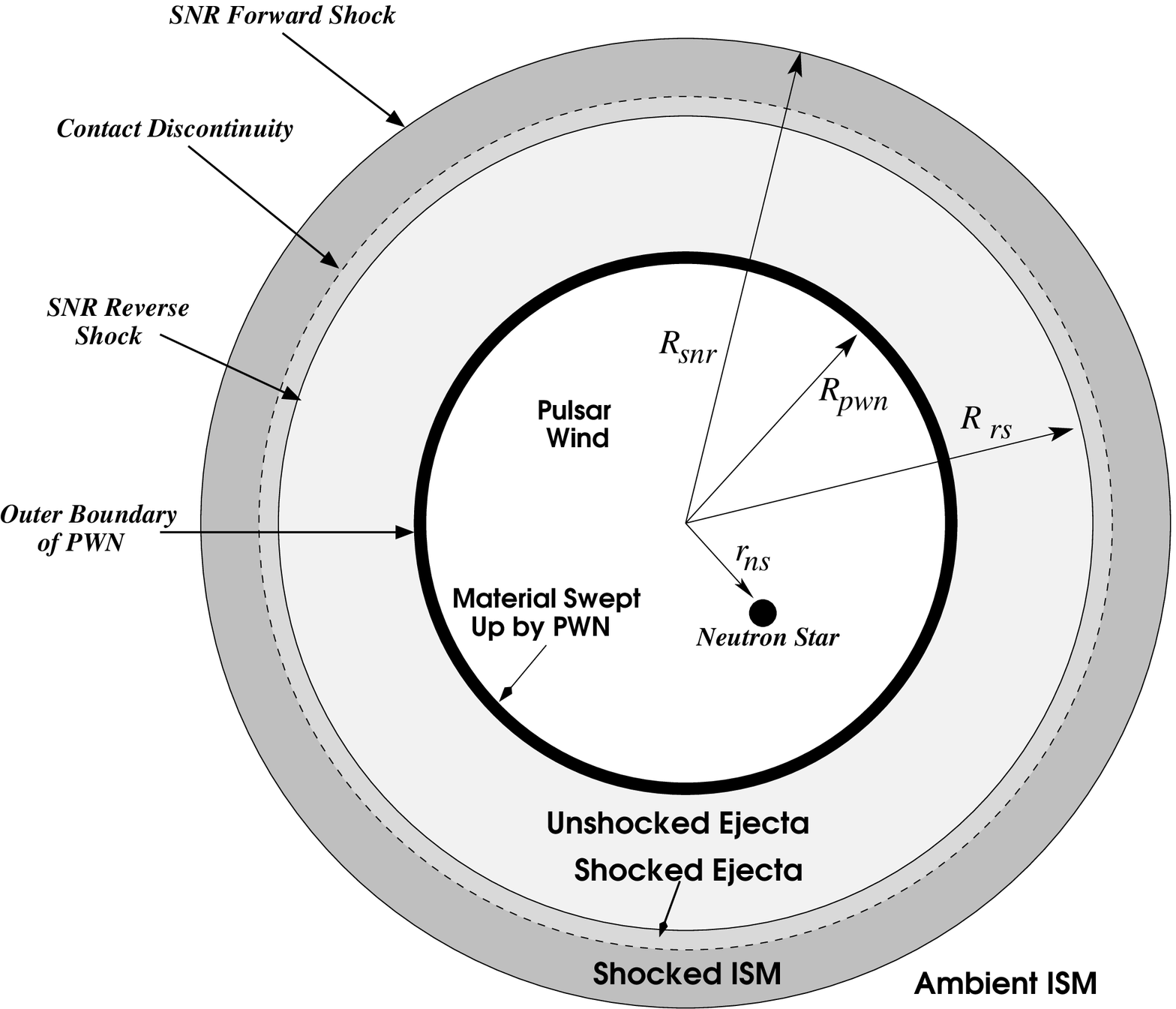}
\end{center}
\figcaption{Diagram of a Composite SNR in the Free Expansion stage of
  its evolution.  In this image, the ratio between the thickness of
  the mass shell surrounding the PWN and the radius of the PWN is
  $1/24$, as determined by \citet{vdswaluw01a}, and the radio of the
  SNR Forward Shock, Contact Discontinuity, and Reverse Shock radii
  are equal to the values given in \citet{chevalier82} for his
  $n=9,s=0$~case. The colors denote the nature of the material within
  each region. \label{snrpwnstruct}} 
\end{figure}

\begin{figure}
\begin{center}
\includegraphics[width=5.00in,angle=0]{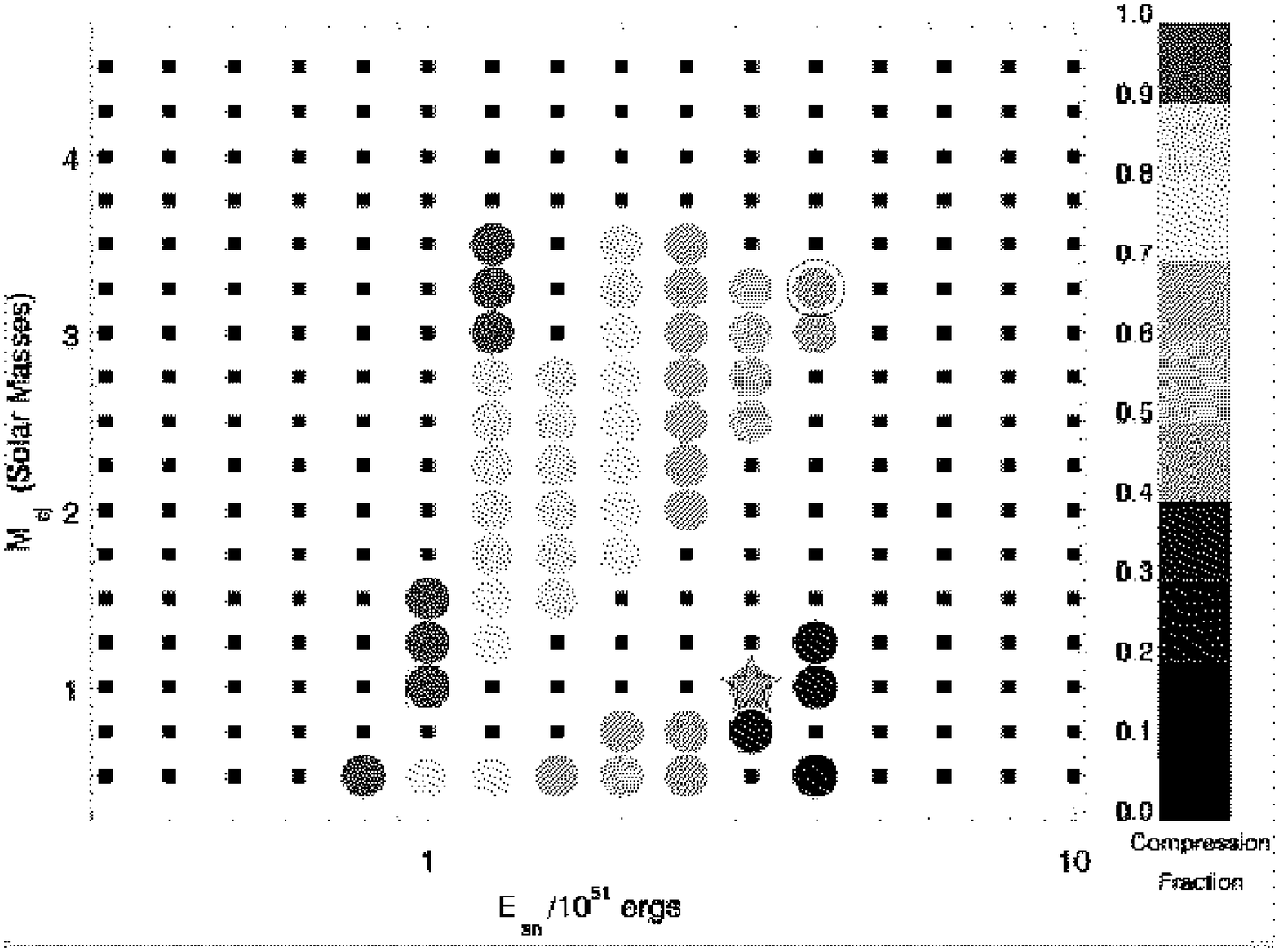}
\includegraphics[width=5.00in,angle=0]{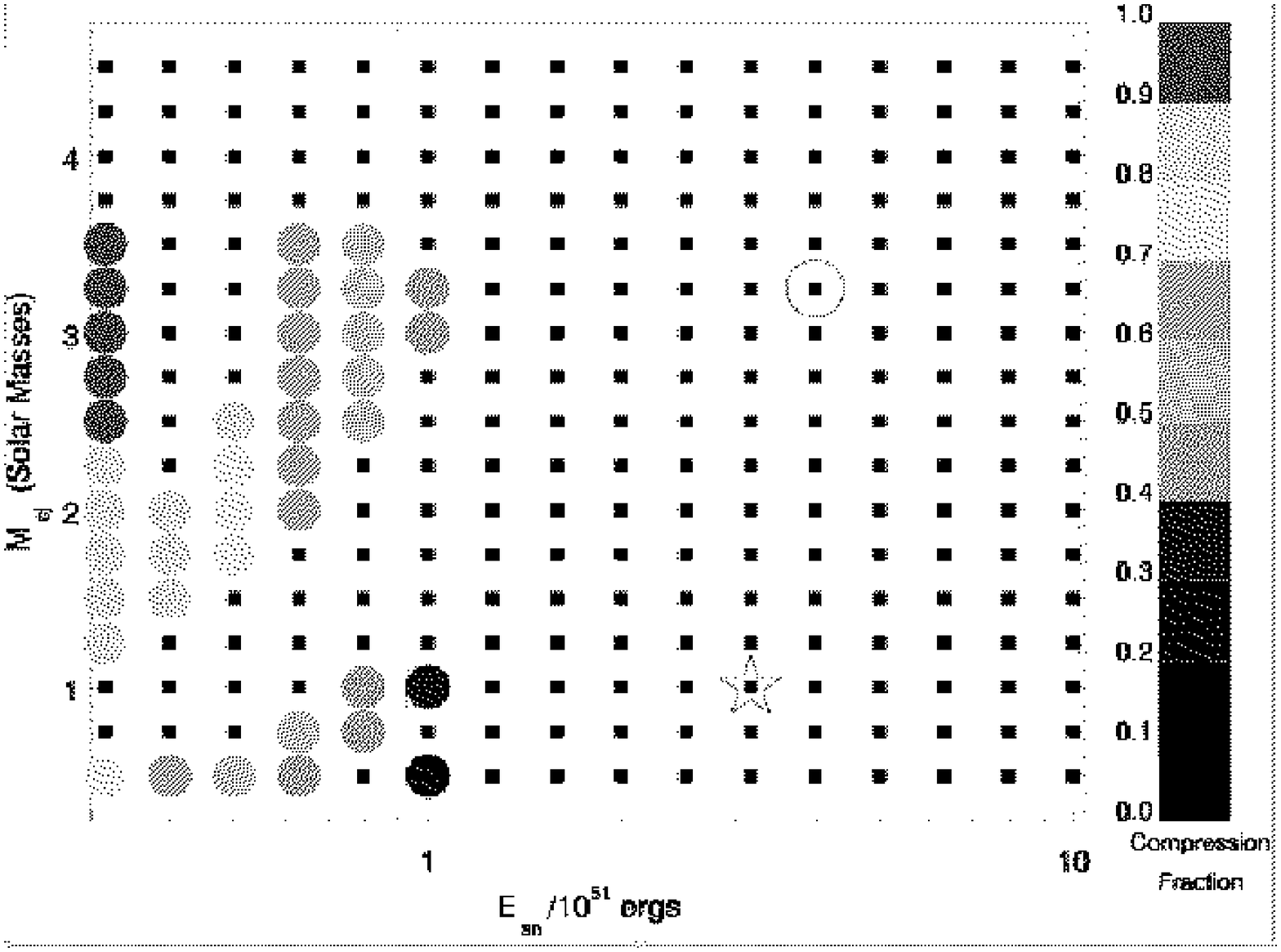}
\end{center}
\figcaption{The results of varying $E_{\rm sn}$ and $M_{\rm ej}$ for a
  $P_0=5$~ms, $B_{\rm ns}=5\times10^{11}$~G ({\it top}) and
  $P_0=10$~ms, $B_{\rm ns}=8\times10^{11}$~G ({\it bottom}) neutron
  star, assuming $n=0.03$~cm$^{-3}$. The black squares indicate which
  scenarios failed the criteria described in \S\ref{g328pwn}, while
  the colored circles indicate those that passed, with the color
  representing the Compression Fraction (defined as the ratio of the
  PWN's volume at the beginning and end of the compression stage) of
  the PWN.  The star indicates a $E_{\rm sn}=1\times10^{51}$~ergs,
  $M_{\rm ej}=1~M_{\odot}$ SN explosion, the square indicates a
  $E_{\rm sn}=3\times10^{51}$~ergs, $M_{\rm ej}=1~M_{\odot}$ SN
  explosion, and the circle indicates a $E_{\rm
  sn}=4\times10^{51}$~ergs, $M_{\rm ej}=3.25~M_{\odot}$ SN
  explosion -- the SN explosion parameters used in
  Fig. \ref{varyns}. \label{varysn}}
\end{figure}

\clearpage
\newpage
\appendix
\section{Equations for the HD Model for the Evolution of a PWN inside
  a SNR}
\label{model2}

In this Appendix, we provide many of the details concerning the
properties of the neutron star, PWN, and SNR needed to simulate
the HD model for the evolution of a PWN inside a SNR described in
\S\ref{model}.  

For the SNR, we assume that the initial ejecta density profile
consists of a constant density core surrounded by a $\rho \propto r^{-9}$
envelope -- the standard assumption for a SNR produced by a Type-II
SNR \citep{chevalier82,blondin01} -- and that the ejecta is expanding 
ballistically ($v_{\rm ej} \equiv r_{\rm ej}/t$).  The boundary
between the constant density core and the outer ejecta envelope has a
velocity $v_{\rm core}$, defined as \citep{blondin01}:
\begin{eqnarray}
\label{vcore}
v_{\rm core} & = & \left(\frac{20}{9}~\frac{E_{\rm sn}}{M_{\rm ej}}
\right)^{1/2} 
\end{eqnarray}
where $E_{\rm sn}$ is the explosion energy of the SN and $M_{\rm ej}$
is the ejecta mass. As a result, the density $\rho_{\rm core}$ of the
ejecta core is \citep{blondin01}: 
\begin{eqnarray}
\label{rhocore}
\rho_{\rm core}(t) & = & \frac{10}{9\pi}~E_{\rm sn}~v_{\rm core}^{-5}~t^{-3}.
\end{eqnarray}
As the SNR expands, it sweeps up and shocks the surrounding
interstellar medium (ISM).  This swept-up material has a higher
pressure than the cold ejecta driving the expansion of the SNR, and as
a result drives a reverse shock (RS) into the SN ejecta.  In between
the outer edge of the SNR, which marks the location of the forward
shock (FS) and the RS is a contact discontinuity which separates the
shocked ISM from the ejecta shocked by the RS.  The pressure inside
the SNR at $r<r_{\rm rs}$, where $r_{\rm rs}$ is the radius of the RS,
is assumed to be zero.  A diagram of this is shown in
Fig.~\ref{snrpwnstruct}.  Since we assume that both the SN ejecta and
the shocked ISM behave as a $\gamma=5/3$ perfect gas, the sound speed
$c_s$ of this material is: 
\begin{eqnarray}
\label{soundspped}
c_s  \equiv \sqrt{\frac{5}{3} \frac{P}{\rho}}
\end{eqnarray}
When the RS is still in the ejecta envelope, we determine the
pressure, velocity, and density profiles on the material between the
RS and FS using the self-similar equations given by
\citet{chevalier82}, evaluating them for the $n=9$, $s=0$ case.
However, when the RS enters the constant density ejecta core, it is no
longer possibly to apply the self-similar solution of
\citet{chevalier82}, and we use the work of \citet{truelove99} to
determine the radius of the RS and the results given by
\citet{bandiera84} to determine the pressure, velocity, and density
profiles between the RS and FS.  It is also necessary to model the
radius of the FS ($R_{\rm snr}$), which we do using the work of
\citet{truelove99}.  This is valid while the SNR is in the Free
Expansion and Sedov-Taylor phases of its evolutions. After the SNR
goes radiative, which occurs at a time $t=t_{\rm rad}$ defined as
\citep{blondin98}:  
\begin{eqnarray}
\label{trad}
t_{\rm rad} & \approx & 2.9 E_{51}^{~\frac{4}{17}} n^{-\frac{9}{17}}
\times 10^4~{\rm yr}.
\end{eqnarray}
After this point, $R_{\rm snr} \propto t^{2/7}$.  An analytic model
for the pressure, velocity, and density distribution of a SNR in this
phase does not currently exist, and therefore it is difficult to
extend our model to this phase, though if one assumes that the
interior of the SNR evolves adiabatically, $R_{\rm pwn}/R_{\rm snr}
\propto t^{0.075}$ for $t>t_{\rm rad}$ \citep{blondin01}.

For the PWN, as mentioned in \S \ref{model}, we assume that it is a
bubble filled with a $\gamma=4/3$ perfect gas.  As a result, the
internal pressure of the PWN, $P_{\rm pwn}$ is equal to: 
\begin{eqnarray}
\label{p_pwn}
P_{\rm pwn} & = & \frac{E_{\rm pwn}}{3 V_{\rm pwn}},
\end{eqnarray}
where $E_{\rm pwn}$ is the internal energy of the PWN and 
$V_{\rm pwn}$ is the volume of the PWN, defined as:
\begin{eqnarray}
\label{v_pwn}
V_{\rm pwn} & = & \frac{4}{3}\pi R_{\rm pwn}^3
\end{eqnarray}
where $R_{\rm pwn}$ is the radius of the PWN.  The internal energy of
the PWN is determined by the rate of energy injected into the PWN by
the neutron star ($\dot{E}$), and energy loss due to its expansion inside the
SNR ($\dot{E_{\rm pwn}^{\rm ad}}$).  For $\dot{E}$, we use the
standard assumption that it is equal to:
\begin{eqnarray}
\label{edoteqn}
\dot{E} & = & \dot{E_0} \left(1+\frac{t}{\tau_0}\right)^{-\frac{p+1}{p-1}}
\end{eqnarray}
where $t$ is the age of the neutron star, $p$ is the pulsar braking
index ($p=3$ for a magnetic dipole), $\tau_0$ is the characteristic timescale 
of pulsar spin-down, and $\dot{E_0}$ is the initial spin-down energy of
the neutron star.  Both $\tau_0$ and $\dot{E_0}$ depend on the physical
properties of the neutron star, with $\tau_0$ defined as
\citep{shapiro83,blondin01}: 
\begin{eqnarray}
\label{taueqn}
\tau_0 & = & \frac{3c^3~I P_0^2}{4\pi^2 B_{\rm ns}^2 R_{\rm ns}^6
  \sin^2 \alpha}  
\end{eqnarray}
where $I$ is the neutron star's moment of inertia,
$P_0$ is the initial spin period, $B_{\rm ns}$ is the magnetic field
of the neutron star, $R_{\rm ns}$ is the radius of the neutron star,
$\alpha$ is the angle between the neutron star's rotation axis and
magnetic field, and \citep{blondin01}: 
\begin{eqnarray}
\label{e0dot}
\dot{E_0} & = & I \left(\frac{2\pi}{P_0} \right)^2 \frac{1}{\tau_0(p-1)}.
\end{eqnarray}
Since the PWN expands adiabatically, $\dot{E_{\rm pwn}^{\rm ad}}$ is
equal to:
\begin{eqnarray}
\label{edotad}
\dot{E_{\rm pwn}^{\rm ad}} & = & -\frac{E_{\rm pwn}}{t}.
\end{eqnarray}
As a result, the change in the internal energy of the PWN over time
($\dot{E_{\rm pwn}}$) is equal to:
\begin{eqnarray}
\label{edotpwn}
\dot{E_{\rm pwn}} & = & -\frac{E_{\rm pwn}}{t} + \dot{E_0}
  \left(1+\frac{t}{\tau_0}\right)^{-2} 
\end{eqnarray}
assuming that $p=3$.  This equation can be solved analytical, and we
result that $E_{\rm pwn}(t)$ can be expressed as:
\begin{eqnarray}
\label{initepwn}
E_{\rm pwn}(t) & = & \dot{E_0}\tau_0
\left(\frac{\ln(1+t/\tau_0)}{t/\tau_0}-\frac{1}{t/\tau_0-1}\right).
\end{eqnarray}
When we run our model, we use Eq. (\ref{initepwn}) to determine the
initial value of $E_{\rm pwn}$, but determine $E_{\rm pwn}$ at later
times using the procedure described in Step 2 in \S \ref{model}.  

During its free-expansion, the PWN is moving faster than its
surroundings, and the mass of the shell surrounding the PWN ($M_{\rm
  sw,pwn}$) is simply: 
\begin{eqnarray}
\label{initmspwn}
M_{\rm sw,pwn}(t) & = & \int_0^{R_{\rm pwn}} 4 \pi R^2 \rho_{\rm ej}(r,t) dr
\end{eqnarray}
where $\rho_{\rm ej}(r)$ is the density profile of the SNR.  After the
collision with the reverse shock, if the PWN is moving faster than its
surroundings we determine the mass of the ejecta shell recently swept
up by the PWN and add it to the value of $M_{\rm sw,pwn}$ calculated
at the time of the reverse shock collision.  If the PWN is moving
slower than its surroundings, we assume that $M_{\rm sw,pwn}$ remains
constant, even if the PWN is being compressed by the surrounding SNR.
Due to the difference in pressure between the PWN interior to the mass
shell and the SNR exterior to the mass shell ($P_{\rm snr}(r=R_{\rm
  pwn})$), the mass shell is subject to a force $F_{\rm \Delta P}$,
defined as:
\begin{eqnarray}
\label{fdelp}
F_{\rm \Delta P} & = & 4\pi R_{\rm pwn}^2 [P_{\rm pwn}-P_{\rm snr}(R_{\rm
    pwn})].
\end{eqnarray}
In this notation, $F_{\rm \Delta P}>0$ means that the PWN interior has
a higher pressure than inside the surrounding SNR.  If the PWN has not
yet encountered the RS, we assume that $P_{\rm snr}(r=R_{\rm
  pwn})=0$.  If the mass shell is moving faster than the sound speed
of the surrounding material ($v_{\rm pwn} > c_s(R_{\rm pwn})$), which
is the case before the PWN interacts with the RS \citep{chevalier92},
the mass shell is decelerated by ram pressure, and the total force on
the mass swept-up by the PWN, $F_{\rm pwn}$, is:
\begin{eqnarray}
\label{forpwn}
F_{\rm pwn} & = & F_{\rm \Delta P}-4\pi R_{\rm pwn}^2
  \rho_{\rm ej}(R_{\rm pwn}) [v_{\rm pwn}-v_{\rm ej}(R_{\rm pwn})]^2.
\end{eqnarray}
If $v_{\rm pwn} < c_s$, then $F_{\rm pwn} = F_{\rm \Delta P}$.  For $t
\ll \tau_0$, analytical solutions to these equations give $R_{\rm pwn}
\propto t^{6/5}$ if the PWN is still inside the central
constant-density core -- a result which is reproduced by our
numerical implementation of the model described in \S \ref{model}.  

In this framework, the period $P$ of a neutron star evolves as:
\begin{eqnarray}
  \label{peqn}
  P & = & P_0\left(1+\frac{t}{\tau_0}\right)^\frac{p-1}{p+1},
\end{eqnarray}
the period-derivative $\dot{P}$ evolves as:
\begin{eqnarray}
  \label{pdoteqn}
  \dot{P} & = &
  \frac{P_0}{2\tau_0}\left(1+\frac{t}{\tau_0}\right)^{-\frac{p-1}{p+1}} 
\end{eqnarray}
and the surface magnetic field $B_{\rm ns}$ of the neutron star,
assuming $p=3$, is: 
\begin{eqnarray}
\label{bnseqn}
B_{\rm ns} & = & 1.5\left(\frac{\dot{E_{0,37}}^\frac{1}{2} P_{0,{\rm
      ms}}^2}{R_{14}^3} \right)\times10^9~{\rm G}  
\end{eqnarray}
where $\dot{E_{0,37}}=\dot{E_0}/10^{37}$~ergs~s$^{-1}$, 
$P_{0,{\rm ms}}$ is the initial period in ms, and $R_{14}$ is the
radius of the neutron star $R/14~{\rm km}$.  Additionally, for $p=3$
$\tau_0$ is equal to: 
\begin{eqnarray}
\label{tauphysical}
\tau_0 & = & 17.3 \frac{I_{45} P_{0,\rm ms}^2}{B_{12}^2 R_{14}^6}~{\rm years}
\end{eqnarray}
where $I=10^{45}I_{45}$~g~cm$^{2}$ and the angle between the spin and
magnetic field axes of the neutron star is $\alpha=45^{\circ}$.
\end{document}